\documentclass[superscriptaddress,amsmath,amssymb,aps,prb,twocolumn]{revtex4-2}

\usepackage{graphicx}
\usepackage{bm}

\usepackage{mathrsfs}

\newcommand{\mycomment}[1]{}

\usepackage{array}
\usepackage{amssymb}
\usepackage{amsmath}
\usepackage{amsfonts}
\usepackage{graphicx,subfigure}
\usepackage{contour}
\usepackage{cancel}
\usepackage{ulem}
\usepackage{gensymb}
\usepackage{bm}
\usepackage{xspace}
\usepackage[colorlinks=true,urlcolor=black,linkcolor=black,citecolor=black]{hyperref}

\usepackage{tikz}
\usetikzlibrary{decorations.pathmorphing,patterns}
\usetikzlibrary{calc,arrows}
\usepackage[compat=1.1.0]{tikz-feynman}
\tikzfeynmanset{/tikzfeynman/warn luatex=false}

\newcommand{\be}{\begin{equation}}
\newcommand{\ee}{\end{equation}}
\newcommand{\bea}{\begin{eqnarray}}
\newcommand{\eea}{\end{eqnarray}}

\newcommand{\f}{\frac}

\newcommand{\Trace}{\mbox{Tr}}

\newcommand{\Qv}{\vec{Q}}
\newcommand{\Rv}{\vec{R}}
\newcommand{\Rvp}{\vec{R}^{\prime}}
\newcommand{\Sv}{\vec{S}}

\newcommand{\lv}{\vec{\ell}}
\newcommand{\lveq}{\vec{\ell}_{\rm eq}}

\newcommand{\cv}{\vec{c}}

\newcommand{\kv}{\vec{k}}

\newcommand{\qv}{\vec{q}}

\newcommand{\qvp}{\vec{q}^{\,\prime}}
\newcommand{\rv}{\vec{r}}
\newcommand{\rvp}{\vec{r}^{\,\prime}}

\newcommand{\uv}{\vec{u}}
\newcommand{\av}{\vec{a}}

\newcommand{\bv}{\vec{b}}

\newcommand{\abslv}{| \lv \, |}

\newcommand{\Jonetwo}{$J_1$-$J_2$\xspace}
\newcommand{\Jeff}[1]{J_{\mathrm{eff},#1}}

\newcommand{\Lambdamat}{\mathbf{\Lambda}}
\newcommand{\Kinvmat}{\mathbf{K^{-1}}}

\makeatletter
\newcommand{\Kinv}{\@ifnextchar\bgroup{\Kinvargs}{K^{-1}}}
\newcommand{\Kinvargs}[3]{K^{-1 #1 #2}_{#3}}
\newcommand{\Dinv}{\@ifnextchar\bgroup{\Dinvargs}{D^{-1}}}
\newcommand{\Dinvargs}[3]{D^{-1 #1 #2}_{#3}}
\newcommand{\D}[3]{D^{ #1 #2}_{#3}}
\newcommand{\Gmaq}[6]{\Gamma^{ #1 \, #2}_{#3 #4, \, #5 #6}}
\newcommand{\Gmaqtilde}[6]{\tilde{\Gamma}^{ #1 \, #2}_{#3 #4, \, #5 #6}}

\newcommand{\gq}{\mathcal G}

\newcommand{\Y}[2]{Y_{#1,#2}}
\newcommand{\X}[2]{X_{#1,#2}}

\newcommand{\normalmodefrequency}[2]{\omega_{#1,#2}}
\makeatother
\newcommand{\OFS}[1]{#1}

\graphicspath{{./}{./Figs/}}

\begin{document}

\title{Nematostriction in frustrated two-dimensional classical Heisenberg models}%
\author{Olav F. Sylju{\aa}sen}
\affiliation{Department of Physics, University of Oslo, P. O. Box 1048 Blindern, N-0316 Oslo, Norway}
\author{Jens Paaske}
\affiliation{Niels Bohr Institute, University of Copenhagen, 2100 Copenhagen, Denmark}
\date{\today}

\begin{abstract}
We investigate the nematic phase transition in the Heisenberg \Jonetwo-model on square and triangular lattices, accounting for finite lattice compressibility and bond-length-dependent magnetic exchange. Using Nematic Bond Theory, a diagrammatic self-consistent method, we study the {\it nematostriction} that happens when the onset of nematic order in the spin-system drives a concomitant structural phase transition. We analyze the mechanisms by which the magnetoelastic couplings renormalize the critical temperature and modify the phonon spectrum. The magnetoelastic \OFS{feedback} can also alter fundamentally the nature of the phase transition. Specifically, on the square lattice, the transition shifts from continuous to weakly first-order (discontinuous) beyond a critical magnetoelastic coupling threshold. Conversely, on the triangular lattice, the transition remains discontinuous regardless of coupling strength.
\end{abstract}

\maketitle

\section{Introduction}\label{sec:Introduction}
Magnetostriction, the elastic deformation of a crystal during a magnetic phase transition, occurs because the exchange coupling between magnetic moments in an insulator depends on the bond length. In two-dimensional frustrated magnets with continuous symmetry, such a structural transition cannot occur due to the absence of magnetic ordering. Nonetheless, vestigial nematic bond ordering could occur at finite temperatures~\cite{Chandra1990, Weber2003Oct}, raising the question if this too is accompanied by a structural phase transition and what the nature of this joint transition is. Such a joint transition has already been established in itinerant electronic systems, for which the influence of a compressible lattice on nematic critical properties is being actively investigated and many interesting and surprising effects have already been uncovered~\cite{Zacharias2015Jul, Karahasanovic2016Feb, Paul2017Jun, Chandra2020, Sarkar2023Dec, Christensen2025Sep}.

Comparatively little is known about how such {\it nematostriction} plays out in local-moment magnets. The classical two-dimensional elastic square-lattice \Jonetwo Heisenberg model was studied in Ref.~\cite{Weber2005} using Monte Carlo simulations at relatively large magnetoelastic couplings, revealing a concomitant nematic and structural phase transition, which was shown to be in the Ising model universality class. Here we revisit the same model in the complimentary small coupling regime, which is likely also the most relevant experimental regime, and consider both a square, and a triangular lattice.

In order to study large system sizes, we employ the Nematic Bond Theory (NBT)~\cite{Schecter2017}, which we extend to also include lattice vibrations. NBT is a self-consistent diagrammatic method for classical spins that treats the local spin-length constraint using a fluctuating Lagrange-multiplier field. It is an extension of the self-consistent Gaussian approximation (SCGA)~\cite{Chalker2017Jan} allowing for spontaneous breaking of the point-group symmetry via the generation of a symmetry breaking momentum-dependent self-energy term. Although NBT is an approximate diagrammatic method neglecting vertex corrections, and is only strictly valid in the limit of infinitely many spin components, $N_{\rm s}\to\infty$, it captures the finite-temperature nematic transition of the \Jonetwo model very well already for the physically relevant case of $N_{\rm s}=3$. This method has the advantage over Monte Carlo simulations that larger system sizes can be reached for modest computing times, and the free energy can be computed directly.  Furthermore, the absence of statistical errors makes it possible to perform accurate extrapolations to the limit of infinite system size.

As in Ref.~\cite{Weber2005}, we find a single joint nematostrictive phase transition. We find evidence, however, that on the square lattice the phase transition becomes discontinuous for magnetoelastic couplings beyond a threshold which is much smaller than the couplings studied in Ref.~\cite{Weber2005}. We map out how the critical temperature is altered by the magnetoelastic couplings, and calculate how the nematic order influences the phonon spectrum. In section~\ref{sec:model} we define the model and establish the generalized NBT equations. In section~\ref{sec:results} we present the results of the numerical investigations, and in section~\ref{sec:disc} we discuss our findings.

\section{Compressible square lattice\\$J_1$-$J_2$ Heisenberg model}\label{sec:model}

The \Jonetwo Heisenberg model with a distance-dependent interaction is
\begin{align}
H_J &= \f{1}{2} \sum_{\rv,\rvp,\alpha} J(\rvp-\rv) S^{\alpha}_{\rv}  S^{\alpha}_{\rvp}, \label{Heisenbergmodel}
\end{align}
where the atomic spins are treated classically as unit vectors. The atom positions are denoted $\rv = \Rv + \uv_{\Rv}$ where $\Rv$ is a site on a reference lattice with lattice vectors $\av_1$ and $\av_2$ containing in total $N=N_x^2$ sites, and $\uv_{\Rv}$ is a displacement.
Expanding about the reference lattice, to first order in the displacements, we find \be
J(\rvp-\rv) \approx   J(\lv \,) + \left( \vec{u}_{\Rv+\lv}- \vec{u}_{\Rv} \right)  \cdot \nabla J(\lv)
\ee
where the vector between interacting spins in the reference lattice is denoted $\lv \equiv \Rvp-\Rv$.
We make the assumption that the exchange couplings $J$ are only functions of the length $|\rvp-\rv|$, and set
$\nabla J(\lv) = g_{\lv} \; \lv/|\lv|$ where $g_{\lv} = \f{\partial J}{\partial r} |_{\abslv}$ which we will refer to as the magnetoelastic coupling.

The displacement vector $\vec{u}_{\Rv}$ is written in terms of the symmetric elastic strain tensor $\epsilon^{ij} \equiv \left( \f{\partial u_i}{\partial r_j}+ \f{\partial u_j}{\partial r_i} \right)/2$, and finite wave vector \OFS{\sout{normal}} phonon modes $\X{m}{\kv}$,
\be
u^{i}_{\Rv} = \epsilon^{ij} R^j + \sum_{\kv \neq 0} W^i_{m,\kv} e^{i \kv \cdot \Rv} \X{m}{\kv}, \label{displacement}
\ee
where repeated latin indices are to be summed over. $\X{m}{\kv}$ is the amplitude of phonon mode $m$ at wave vector $\kv$ and causes a displacement proportional to the normal mode eigenvector $\vec{W}_{\! \! m,\kv}$. Modes with $\kv=0$ are excluded from the second term, as they are accounted for by the (uniform)  strain in the first term.

The energy of lattice deformations is
\be
H_{latt} = \f{1}{2} N c_{ij,kl} \epsilon^{ij} \epsilon^{kl} + \f{1}{2} \sum_{\kv \neq 0} M \omega_{m,\kv}^{2}   | \X{m}{\kv} |^2
\ee
where $c_{ij,kl}$ is the elastic stiffness tensor per site, symmetric in its first and second pair of indices, and $\normalmodefrequency{m}{\kv}$ are the phonon frequencies. $M$ is the atom mass, which we will set to unity.
\mycomment{
It is more common to write the elastic part as
\be
H_{elastic} = \f{1}{2} V \tilde{c}_{ij,kl} \epsilon^{ij} \epsilon^{kl} 
\ee
where $\tilde{c}_{ij,kl}$ is the elastic stiffness tensor per unit volume (energy per unit volume). We use
$c_{ij,kl}$ which the elastic stiffness tensor per site (or per spin). It can be gotten by multiplying the elastic stiffness tensor per area (volume) by the area (volume)
of the unit cell as
\be
V \tilde{c}_{ij,kl} = N c_{ij,kl} \implies c_{ij,kl} = \f{V}{N} \tilde{c}_{ij,kl} = V_u \tilde{c}_{ij,kl}.
\ee
}
The elastic stiffness tensor, phonon frequencies and associated eigenvectors are found by diagonalizing the dynamical matrix
\be
\mathcal{D}_{\kv}^{ij} = \f{1}{N M} \sum_{\rv,\rvp} \f{\partial^2 V_{pot}}{\partial r^i \partial r^{\prime j}} e^{-i \kv \cdot \left( \rv - \rvp \right)} |_{eq}.
\ee
where $V_{pot}$ is the elastic potential energy of the lattice. The summand is evaluated at the atom equilibrium positions $r^i_{eq}= \left( \delta^{ij} + \epsilon^{ij} \right) R^j$ \OFS{($\delta^{ij}$ is the Kronecker-delta)} so as to account for a modified phonon spectrum in the deformed lattice.

We will assume a concrete microscopic model where $V_{pot}$ results from elastic bonds with spring constants
$\alpha_{\lv}$ between sites separated by vectors $\lv$ on the reference lattice, and consider only in-plane displacements.
The dynamical matrix will then be a two-by-two matrix and the number of \OFS{acoustic} phonons branches will be $N_{\rm ph}=2$.
Concretely the expression for the dynamical matrix reduces to
\be
\mathcal{D}_{\kv}^{ij} = \f{1}{M} \sum_{\lv} \alpha_{\lv} \, \f{\ell^i_{\rm{eq}}  \ell^j_{\rm{eq}}}{|\lveq|^2} \left( 1 -  e^{-i \kv \cdot \lveq} \right) \OFS{,} \label{DynamicalMatrix}
\ee
where the sum goes over all neighbors \OFS{that are connected to the atom at the origin of the reference lattice by a spring.} Each of these neighbors has equilibrium coordinates $\ell^i_{\rm{eq}} = \left( \delta^{ij} + \epsilon^{ij} \right) \ell^j$.
For the square lattice we will consider springs with magnitude $\alpha_{1(2)}$ between first (second) neighbors only:
$\alpha_1 \equiv \alpha_{\pm \av_1} = \alpha_{\pm \av_2}$ and $\alpha_2 \equiv \alpha_{\pm (\av_1 - \av_2)} = \alpha_{\pm (\av_1 + \av_2)}$, where $\av_1=(a,0)$ and $\av_2=(0,a)$ and $a$ is the lattice spacing of the reference lattice. While we will generally compute the dynamical matrix in eq.~(\ref{DynamicalMatrix}) taking into account a finite strain $\epsilon^{ij}$, its explicit form for the undeformed ($\epsilon^{ij}=0$) square lattice is
\begin{align}
\mathcal{D}^{xx}_{\kv} &= 2\f{\alpha_1}{M} \left[ 1 - \cos{\left(k_x a \right)}\right] +2 \f{\alpha_2}{M} \left[ 1- \cos{\left(k_x a\right)} \cos{\left(k_y a\right)} \right],
\nonumber \\
\mathcal{D}^{yy}_{\kv} &= 2\f{\alpha_1}{M} \left[ 1 - \cos{\left(k_y a\right)}\right] +2 \f{\alpha_2}{M} \left[ 1- \cos{\left( k_x a \right)} \cos{\left(k_y a \right)} \right],
\nonumber \\
\mathcal{D}^{xy}_{\kv} &= 2 \f{\alpha_2}{M} \sin{\left(k_x a \right)} \sin{\left(k_y a \right)} = \mathcal{D}^{yx}_{\kv}.
\end{align}

The $\kv=0$ limit of $\mathcal{D}_{\kv}^{ij}$ determines the components of the elastic stiffness tensor according to $\mathcal{D}^{ik}_{\kv} \sim \f{1}{M} \sum_{j,l} c_{ij,kl} k^j k^l$,
which gives in the notation of Eq.~(\ref{DynamicalMatrix})
\be
c_{ij,kl} = \sum_{\lv} \alpha_{\lv} \, \f{\ell^i_{\rm eq}  \ell^j_{\rm eq} \ell^k_{\rm eq}  \ell^l_{\rm eq} }{2 |\lveq|^2}.
\ee
This expression is manifestly symmetric in all four indices, which means that all components with the same number of $x$(or $y$)-indices are equal. The non-zero components of the elastic stiffness tensor can conveniently be written as a three-by-three Voigt matrix
\be
V = \begin{pmatrix}
c_{xx,xx} & c_{xx,yy} & c_{xx,xy} \\ c_{yy,xx} & c_{yy,yy} & c_{yy,xy} \\ c_{xy,xx} & c_{xy,yy} & c_{xy,xy}
\end{pmatrix}
\ee
written in the basis $\{ \epsilon^{xx},\epsilon^{yy},\epsilon^{xy}+\epsilon^{yx}\}$. The V matrix can be diagonalized and gives eigenvalues $\mu_n$ and eigenvectors $v_n$ with $n \in \{1,2,3\}$. For the undistorted square lattice these eigenvalues and eigenvectors are
\begin{align}
\mu_1 &= \alpha_1 a^2,   & & v_1= \f{1}{\sqrt{2}} ( 1, -1 ,0)^T, \\
\mu_2 &= \alpha_2 a^2,   & & v_2 = (0,0,1)^T, \\
\mu_3 &= \left(\alpha_1  + 2 \alpha_2 \right) a^2,  & & v_3 = \f{1}{\sqrt{2}} ( 1, 1 ,0)^T.
\end{align}
For concreteness, hereafter we will take $\alpha_1=2\alpha_2 \equiv \alpha$, which corresponds to an isotropic crystal. The corresponding eigenmodes of the strain tensor therefore end up being an orthorhombic mode $\epsilon_{1}=(\epsilon^{xx} -\epsilon^{yy})/\sqrt{2}$ with stiffness $\mu_1$, describing elongation in one direction and compression in the other, a shear mode $\epsilon_2= \epsilon^{xy}+\epsilon^{yx}$, with $\mu_2$ and a volumetric mode $\epsilon_3 = (\epsilon^{xx} + \epsilon^{yy})/\sqrt{2}$ with $\mu_3$ changing the volume (area). We will adhere to this naming of the modes even in the presence of finite strain ($\epsilon^{ij} \neq 0$) as these deformations will turn out to be very small resulting in only a tiny mixing of the modes. In terms of these eigenmodes, the elastic energy takes the following simple form
\begin{align}
\f{1}{2} N c_{ij,kl} \epsilon^{ij} \epsilon^{kl}
= \f{1}{2} N \sum_{n}\mu_{n}\epsilon_{n}^{2}.
\end{align}

Rewriting the spins and couplings in terms of their Fourier transforms 
\mycomment{
We define the following Fourier-pairs
\begin{align}
S^{\alpha}_{\Rv} &= \f{1}{N^{1/2}} \sum_{\qv} S^{\alpha}_{\qv} e^{i \qv \cdot \Rv}, \qquad  S^{\alpha}_{\qv} = \f{1}{N^{1/2}} \sum_{\Rv} S^{\alpha}_{\Rv} e^{-i \qv \cdot \Rv} \\
J^{\alpha \beta}_{\Rv} &= \f{2}{N} \sum_{\qv} J^{\alpha \beta}_{\qv} e^{-i \qv \cdot \Rv}, \qquad J^{\alpha \beta}_{\qv} = \f{1}{2} \sum_{\Rv} J^{\alpha \beta}_{\Rv} e^{i \qv \cdot \Rv}
\end{align}
The volume conventions used in these Fourier transforms corresponds to the choices $\mu=1/2$ and $\gamma=1$ in the previous note.
}
and inserting normal modes and elastic deformations for the displacements, the magnetic Hamiltonian can be written
\begin{align}
H_J=&\sum_{\qv,\qvp,\alpha} S^{\alpha *}_{\qv}\left( J_{\qv} \, \delta_{\qv,\qvp} +  \epsilon_n  \gq_{n,\qv} \, \delta_{\qv,\qvp}\right.\nonumber\\
&\hspace*{22mm}\left.+
i \Gmaqtilde{m}{\qv-\qvp}{}{\qv}{}{\qvp} \X{m}{\qv-\qvp} \right) S^\alpha_{\qvp}\label{HJeq}
\end{align}
where \OFS{$X_{m,\qv-\qvp}$ is the phonon mode amplitude defined in Eq.~(\ref{displacement})}, and $\gq_{n,\qv}$ is the linear combination of Fourier-transformed magnetoelastic couplings that couples to elastic mode $n$ (which has amplitude $\epsilon_n$):
\be
\gq_{n,\qv} = \sum_{\lv} g_{\lv} \, \f{1}{|\lv|} e^{i\qv \cdot \lv}  w_{\lv}^{T}v_{n}
\ee
where the vector $w_{\lv} \equiv (\ell^x \ell^x, \ell^y \ell^y, \ell^x \ell^y)^T$.

Keeping only first and second neighbor magnetoelastic couplings $g_1 \equiv g_{\pm \av_1}=g_{\pm \av_2}$ and $g_2 \equiv g_{\pm (\av_1 + \av_2)} = g_{\pm (\av_1 - \av_2)}$ these are, for the undeformed lattice,
\begin{align}
\gq_{1,\qv} &= \f{g_1 a}{\sqrt{2}} \left[ \cos{\left(q_x a \right)} - \cos{\left( q_y a \right)} \right], \nonumber \\
\gq_{2,\qv} &= - g_2 a \sqrt{2} \sin{\left( q_x a\right)} \sin{\left( q_y a\right)}, \\
\gq_{3,\qv} &= \f{g_1 a}{\sqrt{2}} \left[ \cos{\left( q_x a\right)} + \cos{\left( q_y a\right)} \right] + 2 g_2 a\cos{\left( q_x a \right)} \cos{\left( q_y a \right)}. \nonumber
\end{align}
For the undeformed lattice only $g_1$($g_2$) determines the coupling strength to the first(second) elastic mode, while both determine the coupling to the isotropic volume compression mode.

The spin-phonon interaction vertex is
\begin{align}
\Gmaqtilde{m}{\qv-\qvp}{}{\qv}{}{\qvp} &=  \f{1}{\sqrt{N}} \sum_{\cv} g_{\cv} \, e^{i \qvp \cdot \cv} f_{m,\qv-\qvp,\cv}
\end{align}
with
\begin{align}
f_{m,\qv,\cv} &=\f{1}{2i} \f{c^j}{|\cv|}  W^{j}_{m,\qv} \left( e^{i \qv \cdot \cv} - 1 \right).
\end{align}
Normal modes with zero momenta are not included here, as they correspond to uniform elastic deformations which are explicitly accounted for by $\epsilon$.

The unit length constraint on the spins is taken into account by delta-functions written as an integral over a constraint field $\lambda_{\Rv}$:

\begin{align}
\prod_{\Rv} \delta( | \Sv_{\Rv} | -1 ) &= \int \prod_{\Rv} \f{\beta d \lambda_{\Rv}}{\pi}  e^{- i \beta \lambda_{\Rv} \left( \Sv_{\Rv} \cdot \Sv_{\Rv} -1 \right)}
\end{align}
where we for later convenience have scaled the integration variables by the inverse temperature $\beta$.
The sum in the exponent of the integrand can be written in terms of Fourier transformed quantities
\begin{align}
\sum_{\Rv} \lambda_{\Rv} \left( \Sv_{\Rv} \cdot \Sv_{\Rv} -1 \right) &=
\sum_{\qv \neq 0} \lambda_{\qv}  \left( \sum_{\qvp} \Sv^*_{\qv+\qvp} \cdot \Sv_{\qvp} \right) \nonumber \\
& + \lambda_{\qv=0} \left( \sum_{\qvp} \Sv^*_{\qvp} \cdot \Sv_{\qvp} -N \right)
\end{align}
where for clarity the $\qv=0$ component has been written separately.
\mycomment{
We have defined the Fourier-transform of $\lambda_{\Rv}$ as
\be
\lambda_{\Rv} = \sum_{\qv} \lambda_{\qv} e^{i \qv \cdot \Rv}, \qquad  \lambda_{\qv} = \f{1}{N} \sum_{\Rv} \lambda_{\Rv} e^{-i \qv \cdot \Rv}
\ee
(so $\nu=0$).
}
The quantity $\sum_{\qvp} \Sv^*_{\qv+\qvp} \cdot \Sv_{\qvp}$ can be interpreted as the spatial modulation of the squared spin length with wave vector $\qv$. Thus the integrations over $\lambda_{\qv \neq 0}$ force these modulations to be zero, i.e. no spatial variations of the spin lengths. In contrast the $\lambda_{\qv=0}$ integration forces the sum of the squared spin lengths to add up to $N$. Together they thus enforce the local constraint that each spin has unit length. Within the SCGA, $\lambda_{\qv \neq 0}$ is simply ignored and only $\lambda_{\qv=0}$ is kept. We will include the $\lambda_{\qv \neq 0}$ in an approximate way using diagrams, and emphasize the special role of $\lambda_{\qv=0}$ by writing it as
$\lambda_{\qv=0} = -i \Delta$ where $\Delta$ is a real number.
\mycomment{
i.e.
\be
\lambda_{\qv} = (-i) \Delta \delta_{\qv,0} + \lambda_{\qv} \left( 1- \delta_{\qv,0} \right).
\ee
}

Putting everything together, the partition function becomes
\be
Z = \int D \Delta D\epsilon D\lambda DX DS   \; e^{-S}
\ee
with
\begin{align}
S &= \sum_{\qv,\qvp,\alpha} S^{\alpha *}_{\qv} \left[ \vphantom{ \Gmaqtilde{m}{\qv-\qvp}{}{\qv}{}{\qvp}} \left( J_{\qv} +  \Delta +\epsilon_n \gq_{n,\qv}  \right) \delta_{\qv,\qvp} \right.  \nonumber \\
& \left. \quad \qquad -(-i) \left( \Gmaqtilde{m}{\qv-\qvp}{}{\qv}{}{\qvp} \X{m}{\qv-\qvp} + \lambda_{\qv-\qvp} \right)  \right] S^\alpha_{\qvp} \label{thisaction}    \\
& \quad -\beta N \Delta + \f{\beta N}{2} \mu_n \epsilon_n^2 + \f{\beta}{2} \sum_{\kv,m} M \omega_{m,\kv}^2 |\X{m}{\kv} |^2 ,
\nonumber
\end{align}
where the spins have been rescaled by a factor $\sqrt{\beta}$.
\mycomment{
The integral measures are
\begin{align}
D S &= \prod_{\qv,\alpha} \left( \f{d S^\alpha_{\qv}}{\sqrt{\beta}} \right), \quad
D\Delta = \f{-i \beta \sqrt{N} d\Delta}{\pi}, \nonumber \\
D\lambda & = \prod_{\qv \neq 0} \left( \f{\beta \sqrt{N} d \lambda_{\qv}}{\pi} \right),
\quad D\epsilon = \prod_{n=1}^{3} d \epsilon_n, \nonumber \\
DX      &= \prod_{n=1}^{N_{\rm ph}} \prod_{\qv \neq 0} \left( \f{dX_{\qv,n}}{a} \right),
\end{align}
where $N_{\rm ph}=2$ is the number of phonon modes.
}

The bare inverse spin propagator can be written as a diagonal matrix in $\qv$-space, $\mathbf{K}$, with matrix elements
\be
K_{\qv \qvp} \equiv \left( J_{\qv} + \Delta + \epsilon_{n}  \gq_{n,\qv} \right) \; \delta_{\qv,\qvp}.
\ee

The form of Eq.~(\ref{thisaction}) makes it convenient to define a combined constraint and phonon field $Y_{m,\qv}$ in the following way
\be
\Y{m}{\qv} = \left\{
\begin{array}{ll}
a \lambda_{\qv}/|J_1| &, m =0\\
\X{m}{\qv} &, m \in [1,N_{\rm ph} ]
\end{array}
\right. \label{Yfield}
\ee
where the index $m$ takes integer values starting from 0, so that the constraint field is the zeroth component. $N_{\rm ph}=2$ is the number of phonons modes.
In order to make the dimensions of the field components equal, we have multiplied the constraint field by the ratio of the lattice spacing to the nearest neighbor exchange energy on the undistorted lattice.
With this combined field the coupling to the spins can be written as a matrix $\Lambdamat$ with matrix elements
\be
\Lambda_{\qv \qvp} = \left(-i \right) \Gmaq{m}{\qv-\qvp}{}{\qv}{}{\qvp} \Y{m}{\qv-\qvp}
\ee
where the vertex function is
\begin{align}
\Gmaq{m}{\qv-\qvp}{}{\qv}{}{\qvp} &= \left\{
\begin{array}{ll}
|J_1|/a, & m=0, \\
\Gmaqtilde{m}{\qv-\qvp}{}{\qv}{}{\qvp}  &  m \in [1,N_{\rm ph} ].
\end{array} \right. \label{Vertexfunction}
\end{align}

Generalizing the number of spin components to $N_{\rm s}$ and integrating over the spins we arrive at the following expression for the partition function
\be
Z = C \int  D\Delta D\epsilon DY \; e^{-\left( S_0 + S_2 + S_r \right)}
\ee
where we have factored out the field-independent constants $C$, and divided the remainder into terms according to their powers of $Y$ so that
\mycomment{
so that
\be
C= \left( \sqrt{\f{\pi}{\beta}} \right)^{N N_{\rm s}} \left( \f{\beta \sqrt{N}}{\pi} \right)^{N} |J_1|^{N-1}
\ee
The first factor comes from the integral over the spins, the second and third from the $\lambda$'s and $\Delta$, and the last factor comes from the zeroth compontent of the $Y$ field; the constraint. The measures are
\be
DY = \prod_{\qv \neq 0,m} \left( \f{d\Y{m}{\qv}}{a} \right), \quad D\Delta = d \Delta, \quad D\epsilon = \prod_{n=1}^3 d \epsilon_n,
\ee
}
\begin{align}
S_0 & =-\beta N \Delta + \f{\beta N}{2} \mu_n \epsilon_n^2 + \f{N_{\rm s}}{2} \Trace \ln \mathbf{K}, \\
S_2 &= - \f{N_{\rm s}}{2 \cdot 2} Tr \left( \Kinvmat \Lambdamat \Kinvmat \Lambdamat \right) + \f{\beta}{2} \sum_{\qv, m =1}^{N_{\rm ph}} M \omega_{m,\qv}^2 |\Y{m}{\qv}|^2\nonumber \\
&\equiv \f{1}{2} \mathbf{Y}^\dagger \mathbf{D^{-1}} \mathbf{Y} \\
S_r &= -\sum_{l=3}^\infty \f{N_{\rm s}}{2 \cdot l} \Trace \left( \Kinvmat \Lambdamat \right)^l.
\end{align}
In order to capture symmetry-breaking phenomena going beyond simple perturbation theory, we construct a diagrammatic theory with renormalized propagators.
In particular, $\Kinv$ refers from now on to the spin-spin correlation function with a self-energy addition $\Sigma_{\qv}$ in the denominator
\begin{align}
K^{}_{\qv} &= J_{\qv} + \Delta  + \epsilon_n \gq_{n,\qv} +\Sigma^{}_{\qv} \label{Keq},
\end{align}
and the functional integral over $\mathbf{Y}$ is carried out by doing Gaussian averages with respect to $S_2$ where $\Dinv$ constitutes the renormalized inverse constraint-phonon field propagator,
\begin{align}\label{Deq}
\Dinv{m}{m^\prime}{\qv}&= \beta M \omega_{m,\qv}^2 \; \delta_{m^\prime m} \left( 1 -\delta_{m,0} \right)\\
&+\f{N_{\rm s}}{2} \sum_{\kv}
\Kinv{}{}{\kv} \Gmaq{m}{-\qv}{}{\kv}{}{\kv+\qv}
\Kinv{}{}{\kv+\qv} \Gmaq{m^\prime}{\qv}{}{\kv+\qv}{}{\kv}\nonumber.
\end{align}
With this extension to include the phonons, the self-consistent diagrammatic approximation follows the NBT approach detailed earlier in Refs.~\cite{Schecter2017, Glittum2021}. The spin-spin correlation function is dressed with the Fock self-energy,
\begin{align}\label{Sigmaeq}
\Sigma^{}_{\kv} &= \sum_{\qv} \Gmaq{m}{-\qv}{}{\kv}{}{\kv+\qv} \Kinv{}{}{\kv+\qv} \Gmaq{m^\prime}{\qv}{}{\kv+\qv}{}{\kv} \D{m^\prime}{m}{\qv},
\end{align}
calculated with fully dressed propagators, but leaving out vertex corrections.
Carrying out the functional integral over $\mathbf{Y}$ and rewriting the partition function in terms of renormalized quantities, the partition function becomes\cite{Glittum2021}
\be
Z = C^\prime \int D \Delta D\epsilon \; e^{-S^\prime}
\ee
\mycomment{
  where the constant 
\be
  C^\prime = -i \left( \sqrt{\f{\pi}{\beta}} \right)^{N N_{\rm s}} \left( \f{\beta \sqrt{N}}{\pi} \right)^{N} \left( \f{\sqrt{2\pi}}{a} \right)^{(N-1)(N_{\rm ph}+1)} |J_1|^{N-1}.  \nonumber
\ee
}
where
\begin{align}
S^\prime &= -\beta N  \Delta + \f{\beta N}{2} \mu_n \epsilon_n^2  + \f{N_{\rm s}}{2} \sum_{\qv} \ln{K_{\qv}}
\nonumber \\
& \qquad + \f{1}{2} \sum_{\qv \neq 0} \ln{\det{D^{-1}_{\qv}}}-\f{N_{\rm s}}{2} \sum_{\qv} \left( K^{-1}_{\qv} \Sigma_{\qv} \right). \label{Sprime}
\end{align}
To arrive at this expression, we have neglected diagrams in a systematic large-$N_{\rm s}$ expansion as described in appendix A of Ref.~\cite{Glittum2021}. \OFS{Among these omitted diagrams, the leading order ones are shown in Fig.~\ref{DiagramsSR}, and hence the error in the free energy is ${\cal O}(1/N_{\rm s})$.}
\begin{figure}[t]
\begin{tikzpicture}
\def\radius{0.5cm}
\begin{feynman}
\draw (0,0) circle (\radius);
\foreach \i in {0,...,3}{\vertex (a\i) at (90*\i+45:\radius);}
\diagram*{(a0) -- [photon] (a2);};
\diagram*{(a1) -- [photon] (a3);};
\end{feynman}
\end{tikzpicture}
\hspace{0.5cm}
\begin{tikzpicture}
\def\radius{0.5cm}
\def\distance{1.5cm}
\begin{feynman}
\draw (0,0) circle (\radius);
\draw (\distance,0) circle (\radius);
\foreach \i in {0,...,2}{\vertex (a\i) at ({\radius*cos(-45*\i+45)},{\radius*sin(-45*\i+45)});}
\foreach \i in {0,...,2}{\vertex (b\i) at ({\distance+\radius*cos(45*\i+135)},{\radius*sin(45*\i+135)});}
\foreach \i in {0,...,2}{\diagram*{(a\i) -- [photon] (b\i);};}
\end{feynman}
\end{tikzpicture}
\caption{Leading order omitted diagrams. Wavy lines indicate the constraint-phonon field propagator $D$ which contains a factor $1/N_{\rm s}$. Solid lines indicate the spin propagator $K^{-1}$. A closed solid loop \OFS{carries} a factor $N_{\rm s}$. \OFS{Both diagrams are ${\cal O}(1/N_{\rm s})$.}\label{DiagramsSR}}
\end{figure}

The remaining integrals are performed using the saddle-point method. The saddle-point equations $\f{\partial S}{\partial \Delta} = 0$  and $\f{\partial S}{ \partial \epsilon_n} = 0$ yield
\begin{align}
\beta &= \f{N_{\rm s}}{2N} \sum_{\qv} K^{-1}_{\qv} , \label{betaeq} \\
\beta \epsilon_n &= -\f{N_{\rm s}}{2 \mu_n N} \sum_{\qv} K^{-1}_{\qv} \gq_{n,\qv}\label{epseq}
\end{align}
(no sum over $n$), since the two contributions from the second line of Eq.~(\ref{Sprime}) cancel each other (cf. Ref.~\onlinecite{Glittum2021}).
Eq.~(\ref{betaeq}) enforces the constraint that the average spin length is unity. It should be noted that since $\Sigma_{\qv}$ and $\epsilon_n$ are not fixed quantities, it is possible to have several sets $\{ (\Sigma^{(1)}_{\qv},\epsilon^{(1)}_n,\Delta^{(1)}),(\Sigma^{(2)}_{\qv},\epsilon^{(2)}_n,\Delta^{(2)}),\ldots\}$ which give the same value for the sum in Eq.~(\ref{betaeq}), i.e. the same (inverse) temperature. This is in contrast to the similar saddle-point equation in the SCGA method, where there is no self energy and no elastic term, which results in a unique value of $\Delta$ for a given value of the temperature. Given this possible multiple-valuedness we view Eq.~(\ref{betaeq}) as an equation that gives the temperature for a given value of $\Delta$ (and $\Sigma_{\qv}$, $\epsilon_n$).

We solve the self-consistent equations numerically by iteration.
First, an initial value of $\Delta$, and random values for the self-energy $\Sigma_{\qv}$ are selected. The elastic mode amplitudes $\epsilon_n$ are initially set to zero. Then the first iteration starts by computing the dynamical matrix $\mathcal{D}_{\kv}$ and elastic stiffness tensor $c_{ij,kl}$ which are both diagonalized to give the phonon frequencies $\omega_{m,\kv}$, normal modes $\vec{W}_{m,\kv}$, elastic stiffnesses $\mu_n$ and elastic eigenmodes $v_n$. Then $\Kinv_{\qv}$ is constructed and the inverse temperature $\beta$ and elastic mode amplitudes $\epsilon_n$ are obtained from Eqs.~(\ref{betaeq}) and (\ref{epseq}). Next the constraint-phonon propagator $D^{-1}_{\qv}$ is constructed from Eq.~(\ref{Deq}) and employed to construct a new self-energy $\Sigma_{\kv}$ from Eq.~(\ref{Sigmaeq}). This marks the end of an iteration, and the next iteration starts again by constructing the dynamical matrix and the elastic stiffness using the deformations obtained in the previous iteration.
This is repeated until the temperature obtained in subsequent iterations converge. Specifically we use the convergence criterion that three subsequent iterations are required to have relative temperatures that differ by at most $10^{-10}$. Typically $10-50$ iterations are needed for convergence.

If one carries out this procedure without modifications one will only get convergence to high temperature states. This is because the self-energy will steadily increase in the iterations causing $\Kinv_{\qv}$ to decrease and therefore also the inverse temperature according to Eq.~(\ref{betaeq}). To remedy this we will also renormalize the value of $\Delta$ in the iteration procedure. Specifically, in each iteration, after the step where $\Sigma_{\kv}$ is constructed from Eq.~(\ref{Sigmaeq}) we find its minimal value $\Sigma_{\kv^*}$ and renormalize $\Delta \to \Delta - \Sigma_{\qv^*}$.

After reaching convergence, we calculate the Gibbs free energy per spin at zero external pressure as follows
\begin{align}
\f{G}{N} &= -\Delta +  \f{1}{2} \mu_n \epsilon_n^2 + \f{N_{\rm s}}{2\beta N} \sum_{\qv} \ln{\left(\beta K_{\qv}\right)}\label{Geq}\\
&\hspace{-5mm} + \f{1}{2\beta N} \sum_{\qv} \ln{\left( \f{\det{\left(a^2 D^{-1}_{\qv} \right)}}{\OFS{2N} (\beta J_1)^2} \right)}-\f{N_{\rm s}}{2\beta N} \sum_{\qv}  \left( \Kinv_{\qv} \Sigma_{\qv} \right) \OFS{,}\nonumber
\end{align}
where $\Delta$ denotes the renormalized value, \OFS{and normalization factors contained in $C^\prime$ have been included}. We have omitted terms which take the form of a constant times temperature as they do not contribute to free energy differences of different phases, latent heat or the specific heat.
\mycomment{
The expression including the constants is
\begin{align}
\beta G &= -\beta N  \Delta + \f{\beta N}{2} \mu_n \epsilon_n^2  + \f{N_{\rm s}}{2} \sum_{\qv} \ln{K_{\qv}}
\nonumber \\
& \qquad + \f{1}{2} \sum_{\qv \neq 0} \ln{\det{D^{-1}_{\qv}}}-\f{N_{\rm s}}{2} \sum_{\qv} \left( K^{-1}_{\qv} \Sigma_{\qv} \right) \nonumber \\
& \quad +\f{N N_s }{2} \ln{\beta} - N \ln{\beta} \nonumber \\
& \quad +(N-1) \ln{\f{a^{(N_{\rm ph} +1)}}{|J_1|}} -\f{1}{2} (N-1) (N_{\rm ph}+1) \ln{2\pi} \nonumber \\
& \quad -\f{1}{2} N \ln{N} - \f{N N_s}{2} \ln{\pi} +N \ln{\pi} + \pi/2i \nonumber 
\end{align}
which can be written
\begin{align}
\beta G &= -\beta N  \Delta + \f{\beta N}{2} \mu_n \epsilon_n^2  + \f{N_{\rm s}}{2} \sum_{\qv} \ln{\beta K_{\qv}}
\nonumber \\
& \qquad + \f{1}{2} \sum_{\qv \neq 0} \ln{\left( \f{\det{\left( a^2 D^{-1}_{\qv} \right)}}{(\beta J_1)^2} \right)}-\f{N_{\rm s}}{2} \sum_{\qv} \left( K^{-1}_{\qv} \Sigma_{\qv} \right) \nonumber \\
& \quad -\f{1}{2} N (N_{\rm ph}+1) \ln{2\pi} + \f{1}{2} (N_{\rm ph}+1) \ln{2\pi} -\ln{\beta} \nonumber \\
& \quad -\f{1}{2} N \ln{N} - \f{N N_s}{2} \ln{\pi} +N \ln{\pi} + \pi/2 i\nonumber 
\end{align}
which leads to
\begin{align}
  \f{G}{N} &= -\Delta +  \f{1}{2} \mu_n \epsilon_n^2 + \f{N_{\rm s}}{2\beta N} \sum_{\qv} \ln{\left(\beta K_{\qv}\right)} \nonumber \\
  & \qquad + \f{1}{2 \beta N} \sum_{\qv} \ln{\left( \f{\det{\left( a^2 D^{-1}_{\qv} \right)}}{(\beta J_1)^2} \right)}-\f{N_{\rm s}}{2\beta N} \sum_{\qv} \left( K^{-1}_{\qv} \Sigma_{\qv} \right) \nonumber \\
& \quad -\f{1}{2\beta}\left[\ln{N}
+\left( N_{\rm s} -2 \right)\ln\pi + \left( N_{\rm ph}+1 \right) \ln{2\pi} \right] \nonumber
\end{align}
where terms of ${\cal O}(1/N)$ (including extending the sum to include $\qv=0$) have been neglected.
The last line of constants can also be written
\be
-\f{1}{2\beta}\left[\ln{2N}
+\left( N_{\rm s} -1 \right)\ln\pi + N_{\rm ph}  \ln{2\pi} \right] \nonumber
\ee
which leads to the final expression
\begin{align}
  \f{G}{N} &= -\Delta +  \f{1}{2} \mu_n \epsilon_n^2 + \f{N_{\rm s}}{2\beta N} \sum_{\qv} \ln{\left(\beta K_{\qv}\right)} \nonumber \\
  & \qquad + \f{1}{2 \beta N} \sum_{\qv} \ln{\left( \f{\det{\left( a^2 D^{-1}_{\qv} \right)}}{2N (\beta J_1)^2} \right)}-\f{N_{\rm s}}{2\beta N} \sum_{\qv} \left( K^{-1}_{\qv} \Sigma_{\qv} \right) \nonumber \\
  & \quad -\f{1}{2\beta} \left[
    \left( N_{\rm s} -1 \right)\ln\pi + N_{\rm ph}  \ln{2\pi} \right]. \nonumber
\end{align}
The result in Ref.~\onlinecite{Glittum2021} is recovered by setting $J_1=a=1$, $N_{\rm ph}=\epsilon_i=0$, $N=V$ and changing the sign on the definition of $\Sigma_{\qv}$.
The computer code uses the expression
\begin{align}
  \f{G}{N} &= -\Delta +  \f{1}{2} \mu_n \epsilon_n^2 + \f{N_{\rm s}}{2\beta N} \sum_{\qv} \ln{\left(\beta K_{\qv}\right)} \nonumber \\
  & \qquad + \f{1}{2 \beta N} \sum_{\qv} \ln{\left( \f{\det{\left( D^{-1}_{\qv} \right)}}{\beta^2} \right)}-\f{N_{\rm s}}{2\beta N} \sum_{\qv} \left( K^{-1}_{\qv} \Sigma_{\qv} \right) \nonumber \\
& \quad -\f{1}{2\beta}\left[\ln{N} +\left( N_{\rm s} -2 \right)\ln\pi + \left( N_{\rm ph}+1 \right) \ln{2\pi} \right]. \nonumber
\end{align}
}

\section{Results}\label{sec:results}
\subsection{Square lattice}
For the square lattice, we consider first and second neighbor interactions. The corresponding Fourier-transformed interaction reads
\begin{align}
J_{\qv} &= J_1 \left( \cos{q_x} +\cos{q_y} \right) + 2J_2 \cos{q_x}  \cos{q_y} +\mathrm{const}. \nonumber
\end{align}
where we have chosen units such that the reference lattice spacing $a=1$ and the nearest neighbor ferromagnetic (FM) coupling is fixed to be $J_1=-1$. We have also added a constant to fix the minimal value of $J_{\qv}$ to be zero.

We will focus on the parameter regime $J_2 > 1/2$ for which $J_{\qv}$ is minimal at two inequivalent points in the Brillouin zone; $(\pi,0)$ and $(0,\pi)$ and the constant is $2J_2$. These minima will energetically favor spin modulations $\Sv_{\qv=(\pi,0)}$ or $\Sv_{\qv=(0,\pi)}$ corresponding to either vertical or horizontal spin stripes. The selection of one of these orientations $(\pi,0)$ or $(0,\pi)$ can be viewed as breaking the lattice rotational symmetry. Thus coming from the high-temperature disordered (lattice symmetric) phase the system has a nematic phase transition as the temperature is lowered beyond a critical temperature~\cite{Chandra1990} regardless of the fact that the Mermin-Wagner theorem impedes actual long-ranged magnetic order. This phase transition exists for all values $J_2>1/2$. As $J_{2}$ is reduced towards $1/2$, the phase transition temperature goes to zero as the minima of $J_{\qv}$ become more and more shallow and finally connect the $X$- and $Y$-points on the edge of the Brillouin zone along a line-degeneracy to a new minimum at the $\Gamma$-point for $J_{2}<1/2$ for which there is no longer a phase transition breaking the point-group symmetry.

While we have chosen FM $J_1$ here, results for antiferromagnetic (AF) $J_1$ will be identical to that of FM $J_1$ provided the sign of $g_1$ is also changed. This is a consequence of the bipartiteness of the square lattice with $J_1$-$J_2$ interactions for which the sign of $J_1$ can effectively be changed by inverting the spins on one of the sublattices.

\begin{figure}[t]
\begin{center}
\includegraphics[width=0.95\columnwidth]{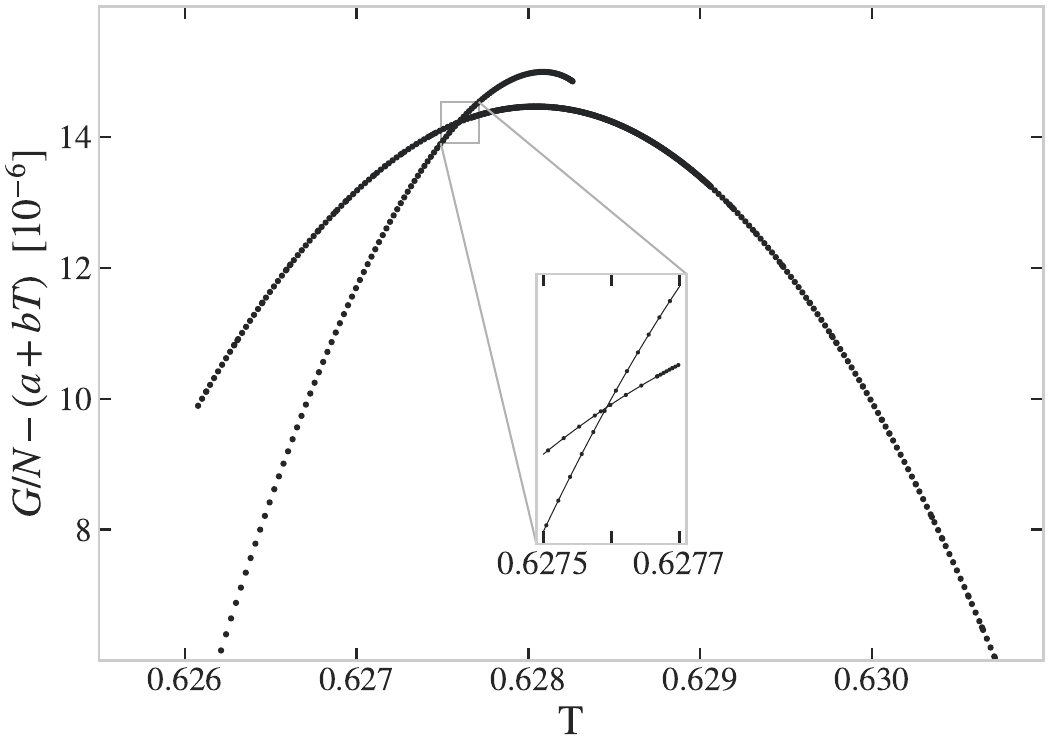}
\caption{Gibbs free energy per site vs. temperature for $J_2=1$ and zero magnetoelastic coupling. $N_x=256$. A linear function $a + bT$ with $a=0.7353$ and $b=-1.65029$ has been subtracted from $G/N$ in order to better visualize the crossing of the two branches at $T_c$.  \label{Fig:nullg}}
\end{center}
\end{figure}

\begin{figure}[t]
\begin{center}
\includegraphics[width=\columnwidth]{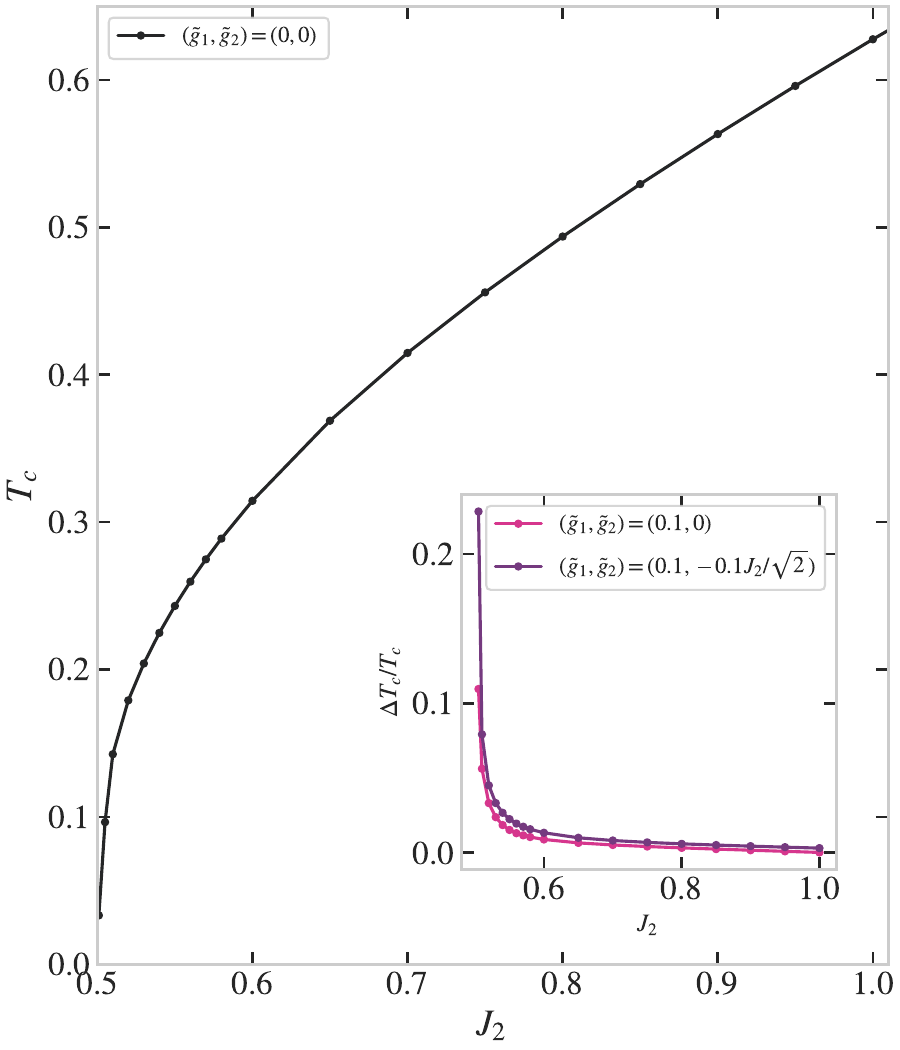}
\caption{Main panel: Critical temperature as a function of $J_2$ for zero magnetoelastic couplings. $N_x=256$.
Inset: Relative change in critical temperatures, $\Delta T_c/T_c \equiv (T_c(\tilde{g})-T_c(0))/T_c(0)$, for two sets of magnetoelastic couplings indicated by the legends.
\label{Fig:tccomparison}
}
\end{center}
\end{figure}

\begin{figure}[t]
\begin{center}
\includegraphics[width=\columnwidth]{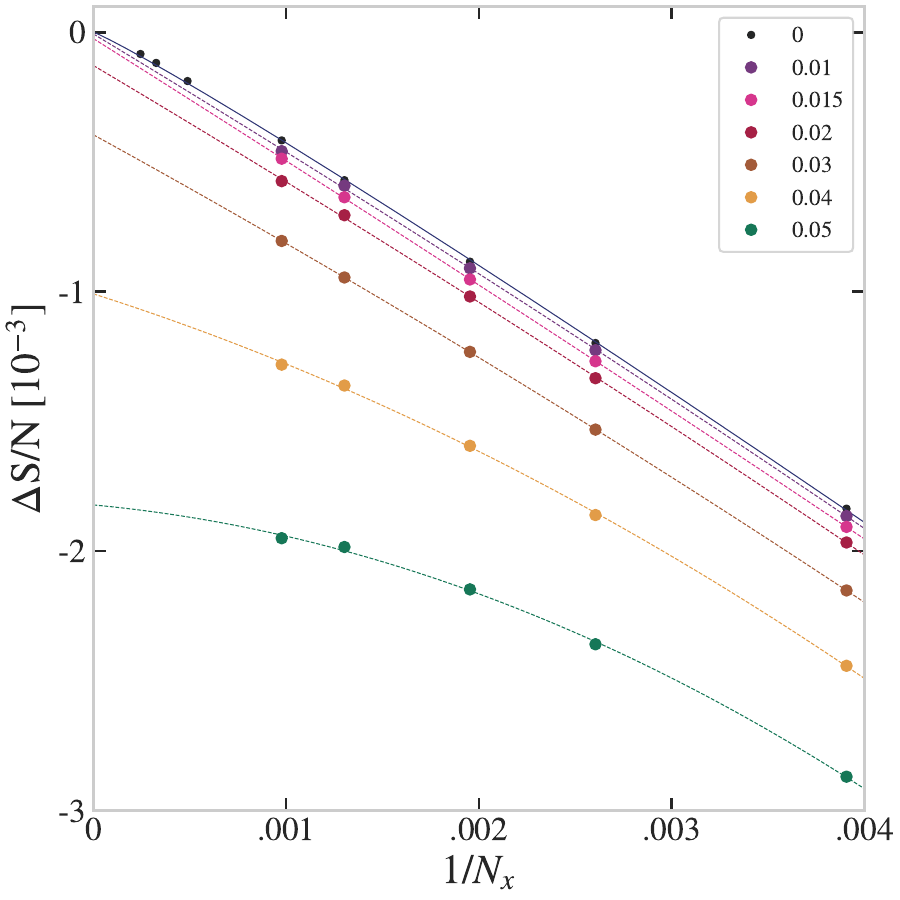}
\caption{Entropy discontinuity per spin $\Delta S/N$ vs. inverse linear system size $1/N_x$ at $J_2=1$. The black small circles show results for no magnetoelastic couplings $(\tilde{g}_i=0)$. The colored circles are for different magnetoelastic couplings $\tilde{g}_1$ as indicated by the legends. $\tilde{g}_2=0$.\label{Fig:disc}
}
\end{center}
\end{figure}

We begin by calculating Gibbs free energy per site for the case where the magnetoelastic couplings are zero and $J_2=1$. The result is shown in Fig.~\ref{Fig:nullg}. It reveals two branches. The low-temperature branch is obtained by starting with a low value of $\Delta$ and first picking a random self-energy. The Eqs.~(\ref{Deq})-(\ref{Sigmaeq}) and  (\ref{betaeq})-(\ref{epseq}) are iterated until convergence, and the temperature and free energy are obtained from Eqs.~(\ref{betaeq}) and (\ref{Geq}) respectively. Then $\Delta$ is increased, but now the converged self-energy and elastic deformations from the previous run is used as initial values. This is repeated to produce the low-temperature branch.
The spin-correlations $(\Kinv_{\qv})$ of converged solutions on the low-temperature branch are peaked on either $\qv=(\pi,0)$ or $\qv=(0,\pi)$ dependent on the initial random self-energy. The low-temperature branch corresponds therefore to the striped phase.
At a certain value of $\Delta$ the low-temperature branch ends abruptly, and the iterations converge to a point on the high-temperature branch at another temperature. Increasing $\Delta$ further traces out the high-temperature branch.
Spin-correlations on the high-temperature branch have equal weights on $(\pi,0)$ and $(0,\pi)$ signifying that the high-temperature branch is the free energy of the disordered (symmetric) phase.
One can also start by decreasing $\Delta$ from a high value. This produces the high-temperature branch that continues until it also ends abruptly and subsequent iterations converge to the low-temperature branch.
The branches cross at the phase transition temperature $T_c$. The discontinuity of slopes at $T_c$ indicates a first order (discontinuous) phase transition with a finite latent heat. However, this is a finite size effect. As will be shown later in Fig.~\ref{Fig:disc}, the discontinuity disappears when $N_x \to \infty$ for zero magnetoelastic couplings.
The crossing points ($T_c$) for other values of $J_2$ are shown in Fig.~\ref{Fig:tccomparison}, black circles. The $T_c$ of these disordered to striped phase transitions is seen to increase with increasing $J_2$ for all $J_2>1/2$, and approaches zero as $J_2 \to 1/2$ consistent with the very shallow minima of $J_{\qv}$ as $J_2 \to 1/2$.

Next, we investigate finite magnetoelastic couplings. In order to scale out the dependence on the magnitude of the spring constants $\alpha$, we transform $X_{m,\kv} \to X_{m,\kv}/\sqrt{\alpha}$ and $\epsilon \to \epsilon/\sqrt{\alpha}$. This causes the system to depend on the magnetoelastic couplings and spring constants as the combinations $\tilde{g}_i \equiv g_i/\sqrt{\alpha}$, which are the variables we will use in the following.
We investigate two sets of magnetoelastic couplings. One with distance-dependent nearest neighbor couplings alone $(\tilde{g}_1,\tilde{g}_2)=(0.1,0)$ and one with both nearest, and second neighbor distance-dependent couplings $(\tilde{g}_1,\tilde{g}_2)=(0.1,-0.1 J_2/\sqrt{2})$. The inset of Fig.~\ref{Fig:tccomparison} shows how the phase transition temperatures deviate from $T_c$ for zero magnetoelastic couplings. As can be seen, these particular values of $\tilde{g}$ give a substantial relative increase in $T_c$ close to $J_2=1/2$ where $T_c$ is already low. For higher values of $J_2$ the relative change in $T_c$ goes rapidly below the one percent level.

\begin{figure}[t]
\begin{center}
\includegraphics[width=\columnwidth]{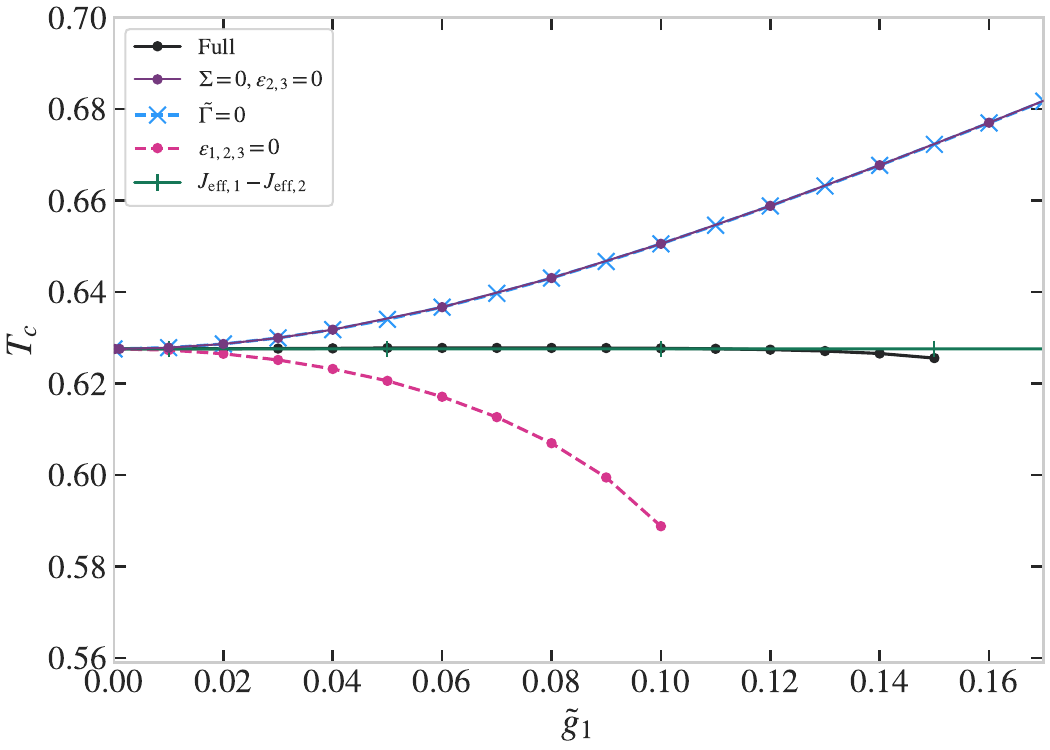}
\includegraphics[width=\columnwidth]{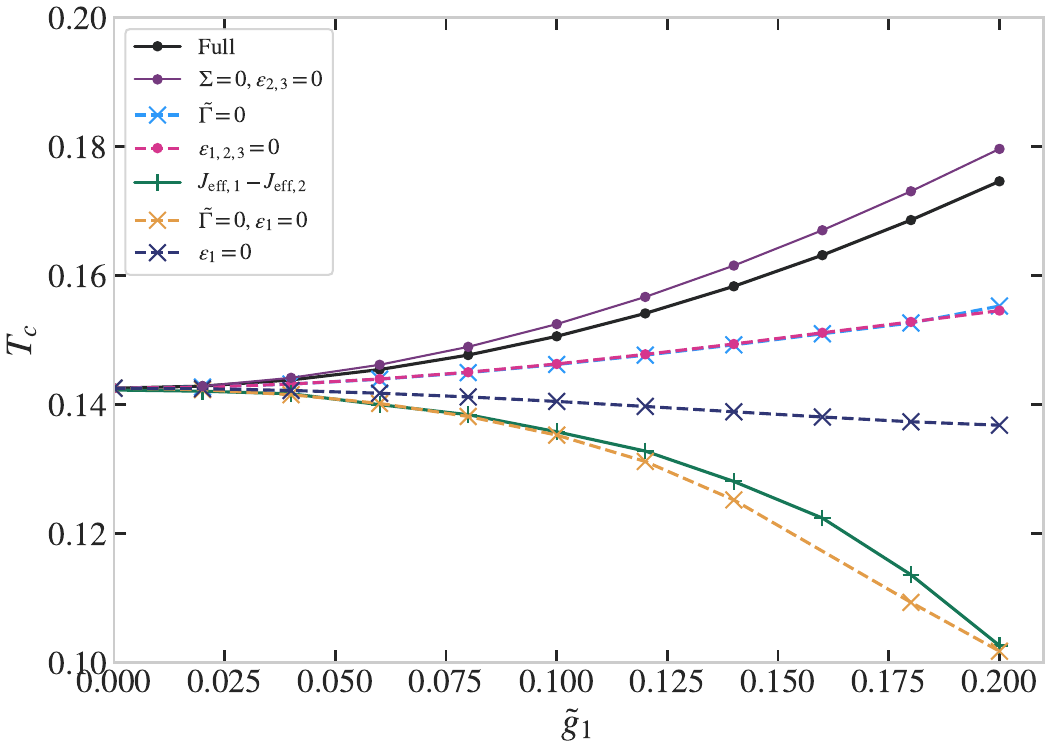}
\caption{Critical temperatures $T_c$ vs. $\tilde{g}_1$ for $J_2=1$ (top panel) and $J_2=0.51$ (bottom panel), with $\tilde{g}_2=0$ and $N_x=256$. The different curves show results obtained under different conditions imposed on the self-consistent equations as indicated by the legends. Full means no extra conditions.
\label{Fig:SQUARE_tcs}}
\end{center}
\end{figure}

In Fig.~\ref{Fig:SQUARE_tcs}, we display the more detailed behavior of the nematic $T_c$ on the strength of the magnetoelastic coupling, focusing on two different parameter values, $J_2=1$ and $J_2=0.51$, for which the inset of Fig.~\ref{Fig:tccomparison} indicates markedly different values of $\Delta T_c/T_c$.
Setting $\tilde{g}_2=0$ we plot in Fig.~\ref{Fig:SQUARE_tcs} $T_c$ as a function of $\tilde{g}_1$ up to the largest possible values for which convergence could be achieved. The black circles show the full result, which reveals a largely $\tilde{g}_1$ independent behavior for $J_2=1$ and a nearly quadratic increase with $\tilde{g}_1$ for $J_2=0.51$.
In order to disentangle the mechanisms that lead to these behaviors, we have solved the self-consistent equations under various simplified conditions, corresponding to the different colored curves in Fig.~\ref{Fig:SQUARE_tcs}.

First, we perform a minimal calculation in which we leave out the self-energy ($\Sigma=0$), corresponding to the SCGA, and switch off both shear and volumetric strain ($\epsilon_{2,3}=0$). The SCGA by itself does not permit breaking of the point-group symmetry, but it is inherently unstable in the sense that any infinitesimal magnetoelastic coupling to orthorhombic strain, $\epsilon_1$, will allow for the symmetry breaking in much the same manner as a finite self-energy. Even without magnetoelastic coupling, the SCGA already encodes the correct value of $T_c$, as the crossover temperature below which the solution of Eq.~\eqref{betaeq}, $\Delta(T)$, vanishes exponentially with temperature and may be interpreted as an inverse squared magnetic correlation length. As observed from the purple dots, this crossover temperature becomes a bonafide nematoelastic critical temperature which increases with $\tilde{g}_1$ in a nearly quadratic manner.

Second, we revert to the full solution, while either leaving out the phonons ($\tilde{\Gamma}=0$, blue crosses) or clamping the system, i.e. fixing all boundary atoms so as to prohibit any uniform elastic deformations ($\epsilon_{1,2,3}=0$, pink dots). For $J_2=1$, leaving out the phonons makes $T_c$ practically identical to the previous (SCGA) case, indicating that the influence of phonons or constraint fluctuations contribute alike to $T_c$. This is not true for $J_2=0.51$, however, where the blue crosses no-longer match the purple. In the latter case, the blue crosses instead match up with the pink dots, implying that leaving out the phonons has the same effect on $T_c$ as clamping the system. For $J_2=1$, on the other hand, the pink dots indicate that phonons cause $T_c$ to decrease with increasing magnetoelastic coupling.

Finally, one may ask how $T_c$ changes with magnetoelastic coupling due to a mere change in the inter-atomic distances deriving from a finite volumetric strain, $\epsilon_3$. For $g_2=0$, only $J_1$ is affected and from Eq.~\eqref{HJeq} one obtains an effective renormalized exchange coupling of
\begin{align}
\Jeff{1} = J_1 + g_1 \epsilon_3/\sqrt{2}.
\end{align}
Extracting the values of $\epsilon_3$ from the fully coupled system at a temperature just above $T_c$ and evaluating the renormalized value $\Jeff{1}$, a corresponding value of $T_c$ can be obtained from the formula known from the pure \Jonetwo-model with no magnetoelastic coupling, $T_c = |J_1| f_{\! \scriptscriptstyle \square}(J_2/|J_1|)$, where $f_{\! \scriptscriptstyle \square}$ is the curve shown in Fig.~\ref{Fig:tccomparison}. The result is shown as the green plusses in Fig.~\ref{Fig:SQUARE_tcs}. For $J_2=1$ this mechanism is seen to have a negligible effect on $T_c$, deriving from the fact that the function $f_{\! \scriptscriptstyle \square}$ is approximately linear, whereby $J_{1}$ cancels from $T_c$. For $J_2=0.51$, on the other hand, the volumetric strain alone leads to a pronounced reduction of $T_c$ with increasing magnetoelastic coupling (cf. green plusses and yellow crosses).

\begin{figure}[t]
\begin{center}
\includegraphics[width=\columnwidth]{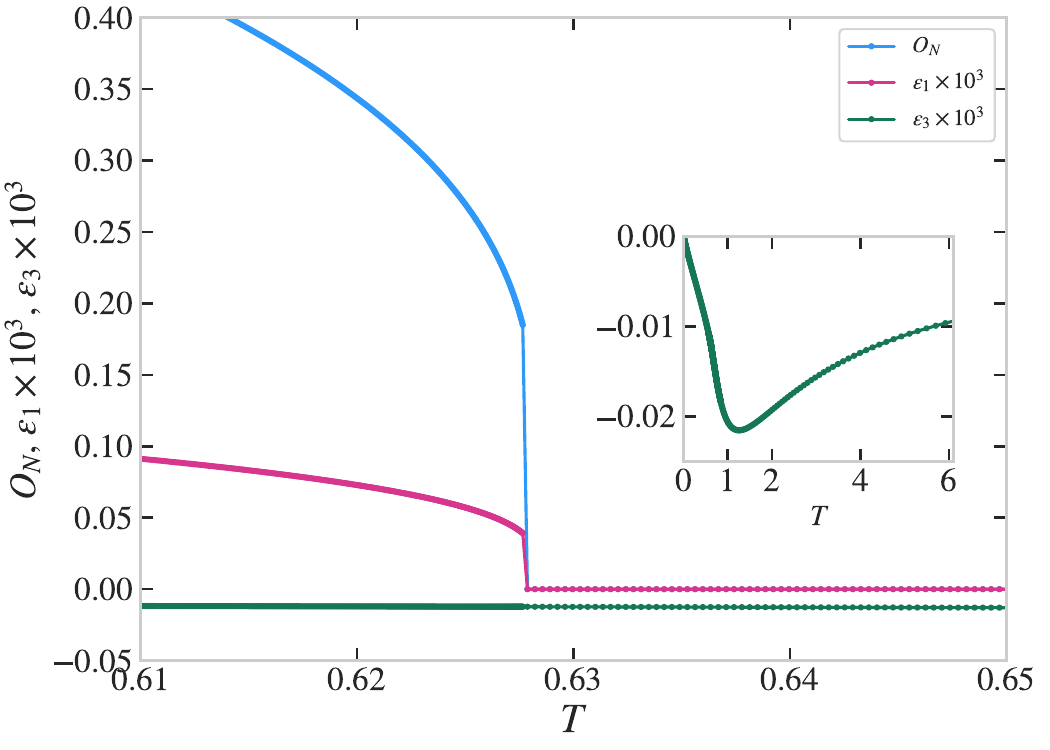}
\caption{Order parameters vs. $T$. Blue curve: Nematic order parameter. Pink curve: Orthorhombic strain $\epsilon_1$. Green curve: Volumetric strain $\epsilon_3$. $\epsilon_1$ and $\epsilon_3$ are multiplied by $10^3$. $(\tilde{g}_1,\tilde{g}_2)=(0.03,0)$, $N_x=256$. The inset shows $\epsilon_3 \times 10^3$ over a wider temperature range.\label{Fig:orderpar}}
\end{center}
\end{figure}
The full equations incorporate all these mechanisms, and we conclude that the almost flat behavior of the black circles for $J_2=1$ is caused by a near cancellation of the increasing $T_c$ due to the orthorhombic strain, $\epsilon_1$, and the decreasing $T_c$ caused by the phonons which takes over for larger values of $\tilde{g}_1$. For $J_2=0.51$, the volumetric strain, $\epsilon_{3}$, alone would lead to a decrease of $T_c$. However, this decrease is almost fully compensated by the phonons to leave $T_c$ nearly constant (dark blue crosses). Its weak dependence on $\tilde{g}_1$ indicates that the effects of the phonons and $\epsilon_3$ on $T_c$ almost cancel, and that therefore the full result (black dots) which includes also the rhombohedral strain, $\epsilon_1$, amounts to a net increase. This explains also the reasonable agreement between the full result (black dots) and the very simplified model with self-energy $\Sigma=0$ and $\epsilon_{2,3}=0$ (pink dots).

The phase transition demonstrates {\it nematostriction} as the system simultaneously distorts and develops lattice nematic order at $T_c$. Fig.~\ref{Fig:orderpar} shows the temperature dependence of both the nematic order parameter $O_N \equiv \f{1}{N} \sum_{\rv} \langle \vec{S}_{\rv} \cdot \left( \vec{S}_{\rv + \hat{x}} - \vec{S}_{\rv+\hat{y}} \right) \rangle$ and the symmetry-breaking orthorhombic strain, $\epsilon_1$, for $\tilde{g}_1=0.03$ near the phase transition. The inset shows the temperature dependence of the volumetric strain, $\epsilon_{3}$, which has a very small discontinuity of the order $10^{-9}$ (cf. green dots in main panel) at  the transition, near which it attains its maximum (largest negative) value, corresponding to a uniform relative reduction of lattice constant of the order of  $10^{-5}$.

As seen in Fig.~\ref{Fig:nullg} the free energy branches approach the crossing point at $T_c$ with distinct slopes. This indicates a first order phase transition with a small discontinuity in entropy. However, this is a finite size effect as can be seen by extracting the entropy discontinuity for several system sizes. The entropy discontinuity $\Delta S \equiv S(T_c^+)-S(T_c^-)$ is obtained by fitting the free energy branches to quadratic polynomials and computing the difference of their negative temperature derivatives at $T_c$.
For $N_x=256$ we find $\Delta S/N = -1.8 \cdot 10^{-3}$, and repeating for other system sizes up to $N_x=4096$ we obtain the values in Fig.~\ref{Fig:disc} shown as small black dots. These values can be fitted to a functional form, $\Delta S = -0.654(1+ 0.05\log{1/N_x} )/N_x$ (solid curve), which tends to zero for $N_x \to \infty$. We conclude from this that the phase transition in the absence of magnetoelastic coupling is continuous in the thermodynamic limit $N_x \to \infty$.

For finite magnetoelastic couplings we also find two branches in the free energy, the crossing of which defines a critical temperature and a discontinuity in entropy, $\Delta S$. The system size dependence of $\Delta S$ at the critical temperature is shown as colored symbols in Fig.~\ref{Fig:disc} for different values of $\tilde{g}_1$ at $J_2=1$. For small values of $\tilde{g}_1$, $\Delta S$ is well fitted by a linear function in $1/N_x$ that extrapolates to very small positive values. This is as for $\tilde{g}_1=0$, black dots, if the logarithmic correction is not taken into account. For finite $\tilde{g}_1$ we are unable to fit the logarithm reliably as system sizes $N_x \gtrsim 1024$ are computationally too demanding when lattice distortions are present. Nevertheless, it is clear that for the largest values of $\tilde{g}_1$, the entropy discontinuity will extrapolate to a finite latent heat in the thermodynamic limit and the phase transition is discontinuous.

\begin{figure}[t]
\begin{center}
\includegraphics[width=\columnwidth]{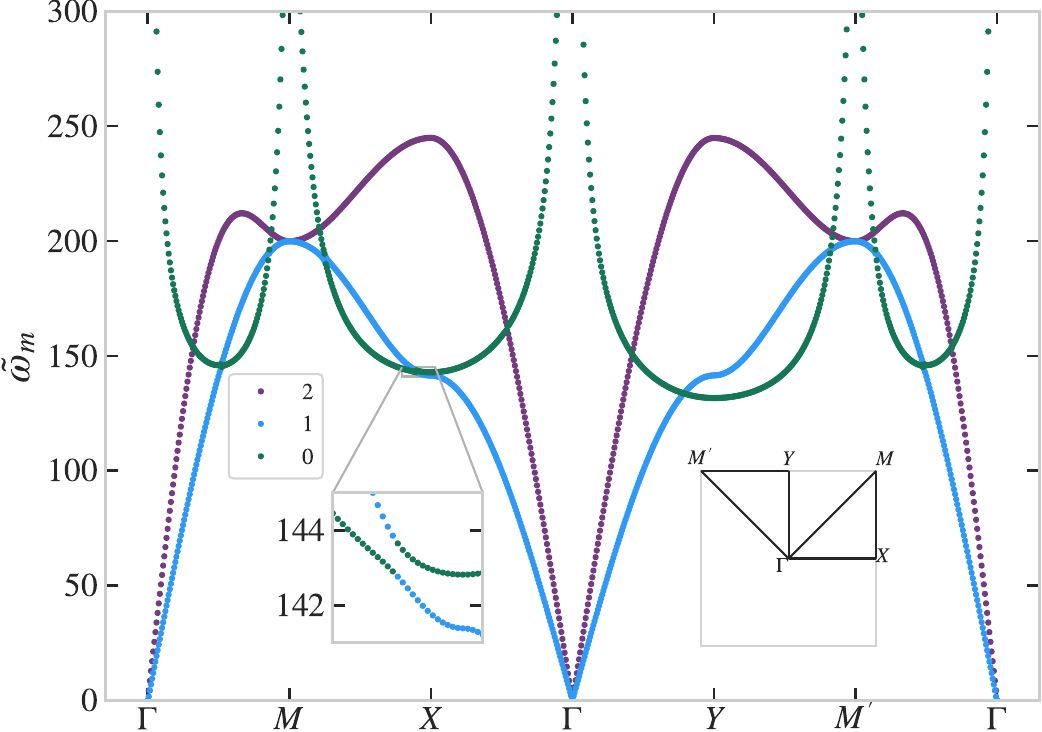}
\caption{The renormalized spectra $\tilde{\omega}_{m,\qv}$ for different $m$ values indicated by the legends. The spectra are obtained at $T=0.604$ just below the phase transition, and is plotted along the Brillouin zone path shown in the right inset. The left inset shows an avoided crossing. $(\tilde{g}_1,\tilde{g}_2)=(0.1,0)$. $J_2=1$, $N_x=256$. \label{Fig:phononscan_square}}
\end{center}
\end{figure}
To estimate the critical value $\tilde{g}_{1c}$ above which the phase transition becomes discontinuous,  we fit our finite size points to a second order polynomial and pick the value of $\tilde{g}_1$ at which it extrapolates to zero.  We note that this procedure of determining $\tilde{g}_{1c}$ strictly gives an upper bound as we cannot rule out the possibility of saturation towards a very small but finite negative value for $N_x > 1024$ for smaller $\tilde{g}$-values.  We find $\tilde{g}_{1c} = 0.01$, see Fig.~\ref{Fig:disc}.
We have also carried out this procedure in the case where the elastic modes are clamped $(\epsilon_{1,2,3}=0)$. We then find practically the same value of $\tilde{g}_{1c}$ \mycomment{$g_c=1.462$}. In contrast, for the somewhat artificial case where there are just elastic modes and no phonons, we get $\tilde{g}_{1c}(\Gamma=0) \approx 0.11$, which is almost an order of magnitude bigger. This implies that the phase transition becomes discontinuous also in the absence of phonons, and that the presence of phonons alter $\tilde{g}_{1c}$.

\begin{figure}[t]
\begin{center}
\includegraphics[width=\columnwidth]{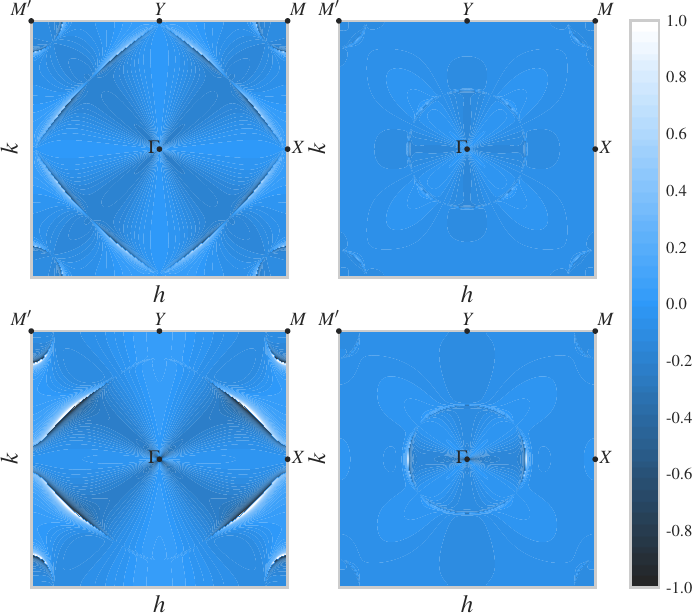}
\caption{Contour plots of the difference between renormalized and bare phonon energies $\tilde{\omega}_{m,\qv} - \omega_{m,\qv}$ for the two phonon branches, $m=1$ (left) and $m=2$ (right). The upper row shows results for $T=0.638$ just above the phase transition, while the lower is for $T=0.604$ just below the phase transition. $(\tilde{g}_1,\tilde{g}_2)=(0.1,0)$, $J_2=1$, $N_x=256$. In order to perform the subtraction when the lattice gets distorted we set $\qv = h \bv_1 + k \bv_2$ and subtract terms with equal values of $(h,k)$. \label{Fig:contourpanels_square}}
\end{center}
\end{figure}

The nematic fluctuations also influence the phonon spectrum. The renormalized phonon spectra $\tilde{\omega}_{m,\qv}$ are obtained as the square root of the eigenvalues of $D^{-1}$ multiplied by $\sqrt{T/M}$. This follows from Eq.~(\ref{Deq}), where the eigenvalues $\tilde{\omega}_{m,\qv}$ are sorted such that the corresponding eigenvector has largest weight on component $m$. This ensures that the renormalized spectra for $m=1,2$ are equal to the bare phonon spectra for zero magnetoelastic coupling. The $m=0$ component, $\tilde{\omega}_{0,\qv}$, is related to the constraint field. In Fig.~\ref{Fig:phononscan_square} we have plotted the renormalized spectra along a path in the Brillouin zone for a temperature just below $T_{c}$. The eigenvalues for the renormalized constraint field (green) exhibit a clear $XY$-asymmetry, while the renormalized phonons (blue and purple) are almost unchanged. However, when plotting the difference between the renormalized and the bare phonon spectra in Fig.~\ref{Fig:contourpanels_square}, one observes marked renormalization both above, and below $T_{c}$. This is seen as smooth softenings together with more pronounced sharp features at $\qv$-points corresponding to the avoided crossings of the $m=0$ component and the other components in Fig.~\ref{Fig:phononscan_square}. Clear signs of XY-anisotropy are seen in the two lower panels of Fig.~\ref{Fig:contourpanels_square}, which are obtained below $T_c$. This rather large anisotropy is caused by the nematic symmetry breaking through the second term in Eq.~\eqref{Deq} and is much larger than the elastic deformations, which in this case leads to new lattice vectors $a_x=1.00024$ and $a_y=0.99971$ for $T=0.604$.

\subsection{Triangular lattice}
\begin{figure}[t]
\begin{center}
\includegraphics[width=\columnwidth]{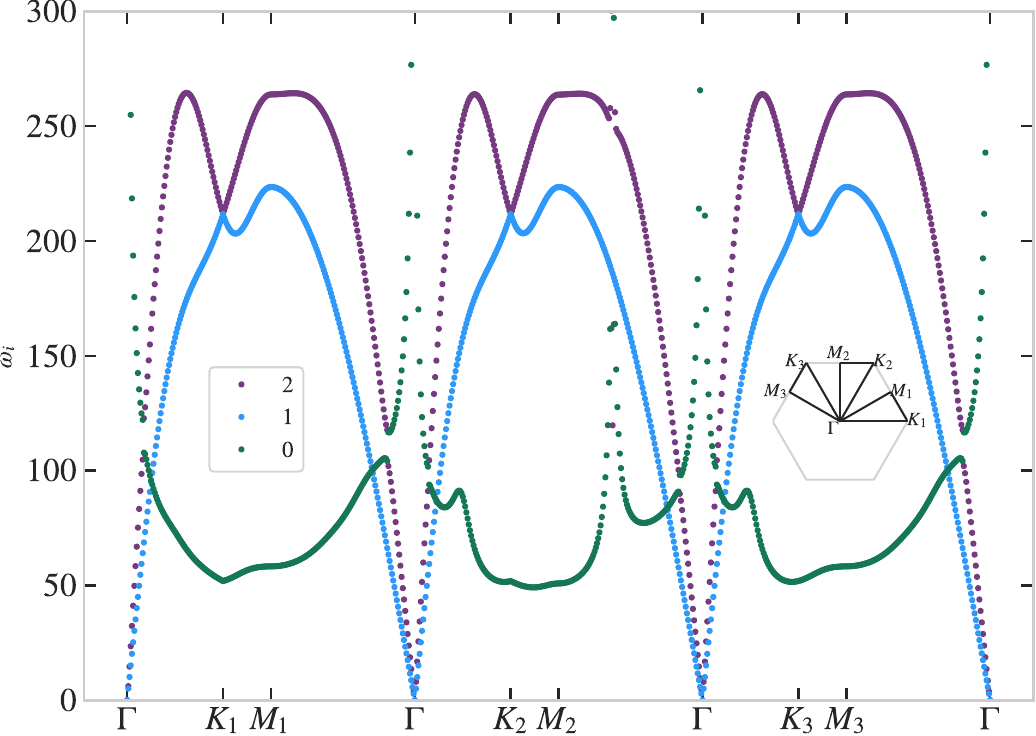}
\caption{The renormalized spectra $\tilde{\omega}_{m,\qv}$ for different $m$ values indicated by the legends. The spectra are obtained for $\tilde{g}_1=0.4$ and $N_x=240$, at $T=0.604$ just below the phase transition, and they are plotted along the Brillouin zone path shown in the right inset. \label{Fig:phononscan_triangular}}
\end{center}
\end{figure}

For the triangular lattice we choose lattice vectors $\av_1 = (1,0)$ and $\av_2 = \left(-1, \sqrt{3}\right)/2$. We have chosen units where the lattice spacing $a=1$. The associated reciprocal lattice vectors are $\bv_1 = \left(1,1/\sqrt{3}\right)2\pi$ and $\bv_2 = (0, 2/\sqrt{3})2\pi$. Defining also $\av_3 \equiv -\av_1 -\av_2$, the notation $q_i \equiv \qv \cdot \av_i$ allows writing the Fourier-transformed exchange coupling as
\begin{align}
J_{\qv} &= J_1 \left[\cos{q_1} + \cos{q_2} + \cos{q_3} \right] \\
&\quad + J_2 \left[\cos{\left( q_1-q_2 \right)} + \cos{\left(q_2-q_3 \right)} + \cos{\left(q_3-q_1\right)} \right] \nonumber \\
&\quad + \mathrm{const}. \nonumber
\end{align}
We consider FM nearest-neighbor exchange, and set $J_1=-1$, and AF next-nearest-neighbor exchange, $J_2>0$. For $J_2 > 1/3$ the minima of $J_{\qv}$ are located on the $\Gamma-\mathrm{M}$ lines in momentum space. These three lines are at a $60\degree$ angle with each other (cf. inset in Fig.~\ref{Fig:phononscan_triangular}), indicating the threefold lattice symmetry which can now be spontaneously broken at low temperatures. The spin pattern in the corresponding magnetically ordered phase at zero temperature has spins aligned along one of the three lattice directions, and spins rotating as one moves perpendicular to this direction.

We will focus on the value $J_2=1/2$ for which the $J_{\qv}$ minima are at $\qv=\Qv$ such that $6\Qv = \vec{G}$ is a reciprocal lattice vector, corresponding to
\be
\Qv \in \pm\f{\pi}{3}\left\{\left(0,\f{2}{\sqrt{3}} \right), \left(1,\f{1}{\sqrt{3}}\right), \left(-1,\f{1}{\sqrt{3}} \right) \right\}.
\ee
Attaching springs with force constants $\alpha$ between nearest neighbors on the triangular lattice gives the dynamical matrix for in-plane phonons. In the case of the undeformed lattice, its components are
\begin{align}
\mathcal{D}^{xx}_{\kv} &= \f{\alpha}{M} \left( 3 - 2\cos{k_x} - \cos{\f{k_x}{2}} \cos{\f{\sqrt{3} k_y}{2}}\right),
\nonumber\\
\mathcal{D}^{yy}_{\kv} &= \f{\alpha}{M} 3\left( 1 - \cos{\f{k_x}{2}} \cos{\f{\sqrt{3} k_y}{2}} \right),\nonumber
\\
\mathcal{D}^{xy}_{\kv} &= \f{\alpha}{M} \sqrt{3} \sin{\frac{k_x}{2}} \sin{\frac{\sqrt{3} k_y}{2}}
= \mathcal{D}^{yx}_{\kv}.
\end{align}
Note that in contrast to  the square lattice, it is not necessary to add next-nearest neighbor springs to ensure stability of the triangular lattice. The elastic modes have, for the undeformed lattice, stiffnesses $\mu_1=3\alpha/4$, $\mu_2=3\alpha/8$ and $\mu_1=3\alpha/2$ with corresponding Fourier-transformed magnetoelastic couplings
\begin{align}
\gq_{1,\qv} &= g_1 \f{1}{\sqrt{2}} \left( \cos{q_x} -  \cos{\f{q_x}{2}} \cos{ \f{\sqrt{3} q_y}{2}} \right),\nonumber\\
\gq_{2,\qv} &= -g_1 \f{\sqrt{3}}{2} \sin{\f{q_x}{2}} \sin{ \f{\sqrt{3} q_y}{2}},\nonumber\\
\gq_{3,\qv} &= g_1 \f{1}{\sqrt{2}} \left( \cos{q_x} + 2\cos{\f{q_x}{2}} \cos{ \f{\sqrt{3} q_y}{2}} \right),
\end{align}
where we have assumed that only the nearest neighbor exchange coupling depends on distance.
\begin{figure}[t]
\begin{center}
\includegraphics[width=\columnwidth]{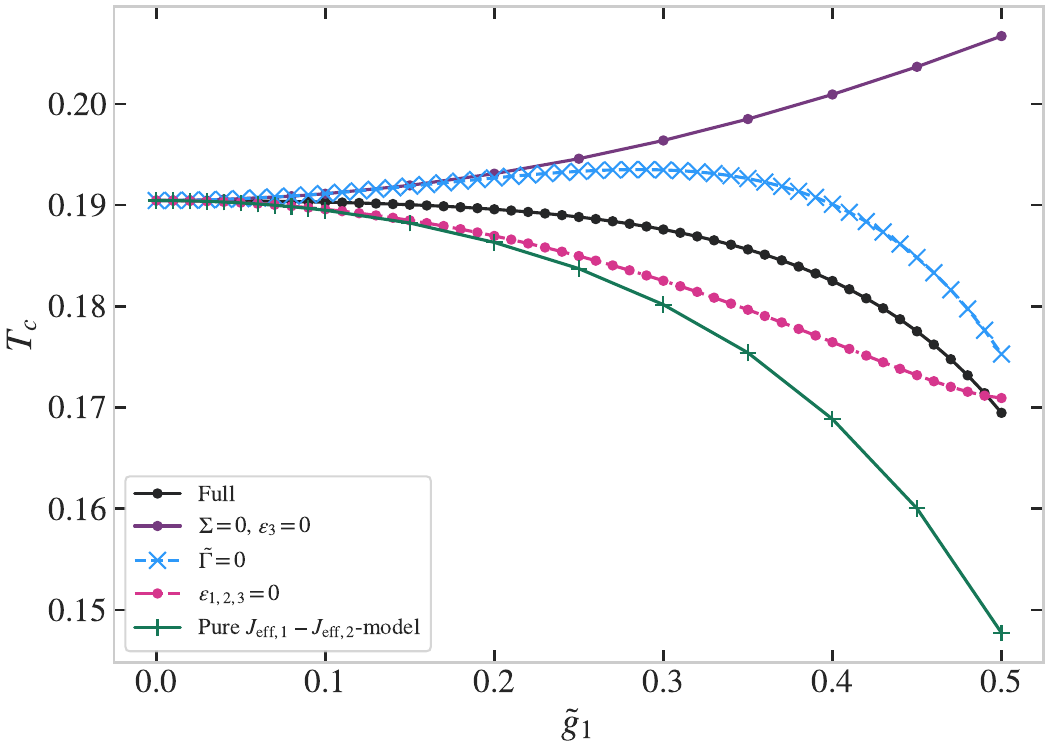}
\caption{Critical temperature $T_c$ vs. $\tilde{g}_1$ for the triangular lattice with $J_1=-1$ and $J_2=0.5$. $L=240$.
The different curves show results obtained under different conditions imposed on the self-consistent equations as indicated by the legends. Full means no extra conditions.}\label{Fig:triangTc}
\end{center}
\end{figure}

For the triangular lattice, already for the system with zero magnetoelastic couplings we find a discontinuous phase transition at $T_c \simeq 0.1905$, with a discontinuity which approaches $\Delta S/N=-0.022$ in the infinite size limit. We find that the first order nature of the phase transition persists also at finite values of $\tilde{g}_1$ with only minute changes in $\Delta S/N$.

As displayed for the square lattice in Fig.~\ref{Fig:SQUARE_tcs}, we show in Fig.~\ref{Fig:triangTc} how $T_c$ changes in the triangular lattice as the strength of the magnetoelastic coupling is increased (black circles). Note that since convergence is better on the triangular lattice for the selected exchange couplings, this plot allows us to explore much larger values than for the square lattice. As for the square lattice, $T_c$ stays almost constant before it clearly decreases for larger values of $\tilde{g}_1$. To investigate this we have repeated our analysis where we solve the self-consistent equations under different simplified conditions. In contrast to the symmetry breaking pattern on the square lattice, which only couples to $\epsilon_1$ and not $\epsilon_2$, the three-fold symmetry breaking on the triangular lattice involves both $\epsilon_1$ and $\epsilon_2$. As for the square lattice, the purple ($\Sigma=\varepsilon_{3}=0$) and the blue (no phonons, $\tilde{\Gamma}=0$) points agree on a quadratic increase up to roughly $\tilde{g}_{1}\approx 0.15$, beyond where they depart rapidly. As for the square lattice, the clamped system (pink circles) exhibits an initial downturn in $T_c$, which however levels off at larger values of $\tilde{g}_{1}$, which were not available for the square lattice. Even though it leads to a decrease in $T_c$, phonons are therefore not able to explain the main downturn of the full solution (black circles). Altogether, this indicates that it is the volumetric strain, $\epsilon_3$, rather than the phonons, which leads to the main decrease in $T_c$ for large $\tilde{g}_1$. To confirm this, we once again extracted the $\epsilon_3$ values for the fully coupled system at a temperature just above $T_c$ and computed the effective exchange coupling as $\Jeff{1} = J_1 + g_1 \epsilon_3/\sqrt{2}$, which was then inserted into the numerically obtained function for $T_c$ from the pure \Jonetwo model, $T_c = |J_1| f_{\! \scriptscriptstyle \triangle}(J_2/|J_1|)$. The resulting curve (green pluses) clearly shows the same trend at the largest values of $\tilde{g}_1$ as the other cases with finite $\epsilon_3$ (black and blue), thus it is reasonable to attribute the main $T_c$ downturn at the largest $\tilde{g}_1$ to the sensitivity of the pure \Jonetwo model towards a stronger value of $J_1$ which arises because of the uniform contraction of the lattice.

Finally we investigate how the nematic ordering affects the phonon spectrum for the triangular lattice. Fig.~\ref{Fig:phononscan_triangular} shows the renormalized spectra along three symmetry equivalent paths in the Brillouin zone that would be identical if the three-fold nematic symmetry was not broken. The figure is obtained for a temperature just below the phase transition and the $m=0$ component clearly reveals a lack of symmetry. As for the square lattice the effect on the phonon spectra is generally weak, except for a very narrow patch on the line from the $M$-points to $\Gamma$ where the $m=2$ phonon mode (purple) is radically softened.  These momentum space locations, like for example $\pm (0,2.27)$, are reasonably close to the momentum vectors, $\pm 2\Qv \approx (0,2.42)$, which connect the two broken symmetry selected minima of $J_{\qv}$.

\begin{figure}[t]
\begin{center}
\includegraphics[width=\columnwidth]{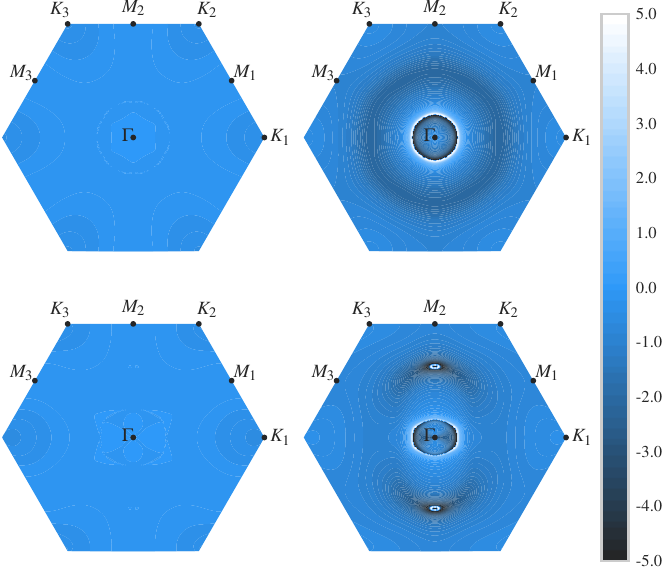}
\caption{Contour plots of the difference between renormalized and bare phonon energies $\tilde{\omega}_{m,\qv} - \omega_{m,\qv}$ for the two phonon branches, $m=1$ (left) and $m=2$ (right). The upper row shows results for $T=0.203$ just above the phase transition, while the lower is for $T=0.172$ just below the phase transition. $N_x=240$. \label{Fig:contourpanels_triangular}}
\end{center}
\end{figure}

In Fig.~\ref{Fig:contourpanels_triangular} we have plotted the difference of the renormalized and bare phonon spectra for the two phonon modes. In order to enlarge other features than the aforementioned large phonon softening close to $2\Qv$ we have only plotted contours in a narrow region about 0. The effects on the phonon mode $1$ is weak, but for phonon mode 2 one can see that the nematic symmetry breaking affects the phonons mainly in two regions encircling the $\Gamma$-point. In the circular region closest to the $\Gamma$-point there is softening of the phonon frequencies just inside and stiffening just outside. For temperatures below $T_c$ the nematic order imprints its asymmetry on the phonon spectra, and one can clearly see large distortions of the circular regions.

\section{Discussion}\label{sec:disc}

We have investigated the effects of a weak magnetoelastic coupling on the finite-temperature nematic phase transition diplayed by the paradigmatic frustrated two-dimensional \Jonetwo Heisenberg model. For both the square, and the triangular lattice, we have found that the coupling to an elastic lattice leads to a nematostrictive phase transition.
We have further analyzed how the magnetoelastic couplings alter the critical temperature of this phase transition, and find a rather complicated picture where several mechanisms together can result in an increase or a decrease of $T_c$ dependent on the specific microscopic details of the system.

For the triangular lattice the transition was found to be discontinuous, regardless of the strength of the magnetoelastic coupling at least up to $\tilde{g}_{1}=0.5$. In contrast, our extrapolations to infinite system size for the square lattice indicate that the transition is continuous for $\tilde{g}_{1}<0.01$, and discontinuous for larger couplings at least up to the largest converged value of $\tilde{g}_{1}=0.15$.
Leaving out the phonons, was found to increase the continuous regime, but still resulted in a discontinuous phase transition, hinting that phonons are not essential for understanding the source of the transition becoming discontinuous.

The phase transition results reported here provide examples of nematostrictive scenarios. The extent to which the details also apply to other parameter values is unclear. In future works it would be interesting to extend our treatment to other values of $J_2$, and also to AF $J_1$, especially to parameter values where the nematic symmetry-breaking pattern is different\cite{Rastelli1979,Rastelli1980}. Adding a third-neighbor coupling $J_3=J_2/2$ is particularly interesting as it allows the study of nematostrictive ordering of classical spiral spin liquids\cite{Seabra2016,Glittum2021,Glittum2026}.

Our results appear to differ from the Monte Carlo results of Ref.~\cite{Weber2005}, which were found to be consistent with a continuous transition. Nevertheless, the magnetoelastic couplings considered there, $\tilde{g}_{1}\sim 0.5-1.5$, were much larger than what we have managed to get convergence for here \OFS{(}for the square lattice\OFS{)}.
The two results are therefore not necessarily inconsistent, but would appear to indicate an intermediate coupling regime for which the nematostrictive transition is discontinuous.

The fact that a magnetoelastic coupling may cause an otherwise continuous phase transition to become discontinuous was suggested already in Refs.~\onlinecite{Rice1954, Domb1956} and later reassessed in more realistic models~\cite{LarkinPikin1969, Levanyuk1970, Sak1974Nov, Wegner1974Jun, Bergman1976Mar, Pikin1993Mar, Chandra2020, Sarkar2023Dec}. In essence, the discontinuous nature of the transition arises from a sufficiently large discontinuity in the specific heat at the otherwise continuous transition for the clamped system. Since the NBT approach does not provide an effective theory for the local nematic order parameter, we cannot retrace the argument made in Ref.~\onlinecite{LarkinPikin1969}. Nevertheless, one may still gain some insight by considering the effective Landau theory for the homogeneous nematic order parameter. This simplistic mean-field approach, leaving out phonons altogether, is similar to that taken in Ref.~\onlinecite{Bean1962Apr} to show that ferromagnetic ordering may become discontinuous on a compressible lattice.

From a Landau theory perspective, the $C_{4v}$ symmetry of the square lattice allows for a nematoelastic Gibbs free energy per lattice site at zero pressure of the form
\begin{align}
G/N &=a(T-T_{c,0})O_{N}^2 + u_{4}\,O_{N}^4-\Delta\nonumber\\
&+\frac{1}{2}\mu_{n}\epsilon_{n}^{2}+\f{N_{\rm s}}{2N\beta}\sum_{\qv}\ln\left[\beta(J_{\qv}+\Delta+\gq_{3,\qv} \,\epsilon_{3})\right] \nonumber\\
&+\eta\epsilon_{3}O_{N}^{2}+\lambda\epsilon_{1}O_{N}+\zeta \epsilon_{1}O_{N}^{3},\label{eq:Glandau}
\end{align}
including three unspecified symmetry-allowed couplings, $\eta$, $\lambda$ and $\zeta$, which all vanish for vanishing magnetoelastic couplings, $\tilde{g}_{1,2}$.
Leaving out the less important shear mode, $\epsilon_{2}$, the free energy is minimized by
\begin{align}
\epsilon_{1}&=-(\lambda/\mu_{1})O_{N}-(\zeta/\mu_{1})O_{N}^{3},\label{eq:eps1min}\\
\epsilon_{3}&=-(\eta/\mu_{3})O_{N}^{2}+G_{0}'/\mu_{3},\label{eq:eps3min}
\end{align}
with the first derivative with respect to $\epsilon_{3}$ of the exchange part of the free energy (cf. also Eq.~\eqref{epseq}),
\begin{align}
G_{0}' &=-\f{N_{\rm s}}{2N\beta}\sum_{\qv}\f{\gq_{3,\qv}}{J_{\qv}+\Delta},
\end{align}
accounting for the {\it exchange magnetostriction}, i.e. the finite volumetric strain by which the system lowers its total exchange energy, even in the absence of long-range magnetic and nematic order in our two-dimensional system~\cite{Yacovitch1979Mar, Callen1965Jul}. Disregarding the nematic order by setting $O_{N}=0$ in Eq.~\eqref{eq:eps3min}, this volumetric strain is given by $\epsilon_{3}\approx G_{0}'/\mu_{3}$, which is negative and therefore corresponds to an isotropic compression of the crystal. This is consistent with the green line in Fig.~\ref{Fig:orderpar} and already the SCGA ($\Sigma_{\qv}=0$) captures very well the non-monotonous temperature dependence displayed in the inset, taking its largest absolute value near $T_{c}$ and vanishing as $T$ and $T^{-1}$, respectively, for low and high temperatures.

Using Eqs.~\eqref{eq:eps1min} and~\eqref{eq:eps3min} to eliminate the strain from Eq.~\eqref{eq:Glandau}, one arrives at the following effective nematic Landau theory
\begin{align}
G_{\rm min}/N &\approx-\Delta-(G_{0}')^{2}/(2\mu_{3})\nonumber\\
&\hspace{-5mm}+\left[a(T-T_{c,0})-\lambda^{2}/(2\mu_{1})+\eta G_{0}'/\mu_3\right]O_{N}^2\nonumber\\
&\hspace{-5mm}+ \left[u_{4}-\lambda\zeta/\mu_{1}-\eta^{2}/(2\mu_{3})\right]O_{N}^4,
\label{Gmin}
\end{align}
when retaining at most the quartic term. For small enough couplings, the coefficient of the quartic term in Eq.~\eqref{Gmin}, $u_{4}-\lambda\zeta/\mu_{1}-\eta^{2}/(2\mu_{3})$, is positive and the nematostrictive transition remains continuous albeit with a renormalized critical temperature given by
\begin{align}\label{eq:Tc2nd}
T_{c}=T_{c,0}+\lambda^{2}/(2a\mu_{1})-\eta G_{0}'/(a\mu_3).
\end{align}
For large enough couplings, the coefficient of the quartic term may become negative,
\begin{align}\label{eq:disccrit}
u_{4}-\lambda\zeta/\mu_{1}-\eta^{2}/(2\mu_{3})<0,
\end{align}
unless $\lambda\zeta$ becomes negative and overcomes the reduction from the $\eta^{2}$ term. This would imply a discontinuous nematostrictive transition, given that a term of order $O_{N}^6$ will be present with a positive coefficient to stabilize the system. Increasing the magnetoelastic coupling further, the couplings $\eta$, $\lambda$ and $\zeta$ may depend on $\tilde{g}_{1,2}$ in a non-linear manner, which might break with the criterion \eqref{eq:disccrit} and cause the transition to revert back to the continuous nature found in Ref.~\cite{Weber2005}. This remains speculative, however, since our microscopic NBT calculations do not converge when the magnetoelastic coupling becomes too large.

Although the effects of $Y_{m, \vec{q}}$, i.e. phonons and local constraints, are completely left out, these simple Landau theory considerations for a homogeneous order parameter including strain provide a plausible scenario for why the nematostrictive transition on a square lattice becomes discontinuous above a certain magnitude of the magnetoelastic coupling. Within this simplified description, the specific heat discontinuity for the nematic transition in the clamped system is simply $\Delta C_V=a^2 T_{c,0}/(2u_{4})$. Coupling only to the volumetric strain ($\lambda=\zeta=0$), the criterion \eqref{eq:disccrit} may therefore be formulated as
\begin{align}
\mu_{3}<\frac{\Delta C_{V}}{T_{c,0}}\frac{\eta^{2}}{a^{2}}=\frac{2\Delta C_{V}}{T_{c,0}}\left(\frac{\partial T^{\ast}_{c,0}}{\partial\ln V}\right)^{2},\label{eq:disc2}
\end{align}
expressed in physical terms via the volume (i.e. $\epsilon_{3}$) dependent critical temperature, $T^{\ast}_{c,0}(\epsilon_{3})=T^{\ast}_{c,0}-(\eta/a)\epsilon_{3}$, inferred from Eq.~\eqref{eq:Glandau}.
This criterion is consistent with Ref.~\onlinecite{LarkinPikin1969} (cf. also Refs.~\onlinecite{Pikin1993Mar, Chandra2020}), except that there the system was three-dimensional and phonons were included in the analysis, causing $\mu_{3}$ to be replaced by a different combination of bulk and shear modulus, which diverges with vanishing shear modulus.

Considering the bilinear coupling, $\lambda$, by itself ($\eta=\zeta=0$), the quadratic terms in Eq.~\eqref{eq:Glandau} are minimized by $O_{N}=-(\lambda/2a(T-T_{c,0}))\epsilon_{1}$, which implies a softening of the orthorhombic stiffness to $\mu_{1}-\lambda^{2}/(2a(T-T_{c,0}))$, reaching zero and signaling a joint nematostrictive transition at $T=T_{c,0}+\lambda^{2}/(2a\mu_{1})$. This scenario is known to impede the nematic fluctuations and lower the critical dimension~\cite{Levanyuk1970, Karahasanovic2016Feb, Paul2017Jun}, leading to a smaller discontinuity in the specific heat. When this mechanism dominates the effects of $\eta$, one might therefore expect the resulting $\Delta C_{V}$ to become too small for the criterion~\eqref{eq:disc2} to be satisfied, resulting in a continuous transition.

As our numerical NBT results suggest, the magnetoelastic coupling to both volumetric and orthorhombic strain, including the corresponding phonon modes, leads to a complex competition between different effects. Even at the level of Landau theory, these joint effects pose an interesting question for a full renormalization group analysis along the lines of Refs.~\onlinecite{Bergman1976Mar, Karahasanovic2016Feb}. Likewise, it should be interesting to study this competition in the quantum critical scenarios studied in Refs.~\onlinecite{Chandra2020, Sarkar2023Dec}.

\section{acknowledgments}
We acknowledge useful discussions with Morten Holm Christensen. O.F.S thanks Niels Bohr Institutets Fond for financial support, and the Condensed Matter Theory group at the Niels Bohr Institute for hospitality. The computations were performed on resources provided by Sigma2 - the National Infrastructure for High Performance Computing and Data Storage in Norway, and on the Fox supercomputer at the University of Oslo.

\section{Data availability}
The numerical data are openly available\cite{data}.

\bibliography{nemstric}

\end{document}